\documentclass[prd,aps,showpacs,preprintnumbers,amssymb]{revtex4}

\usepackage{epsf}

\def\e3p{$\eta \rightarrow 3 \pi$}

\begin{document}

%\title{
\hfill{\normalsize\vbox{%
\hbox{\rm SU-4252-768}
\hbox{\rm JLAB-THY-02-53}
 }}
\title{Effects of light scalar mesons in \e3p decay}

\author{Abdou Abdel-Rehim$^{\it \bf a}$~\footnote[1]{Email: 
abdou@physics.syr.edu}},

\author{Deirdre Black$^{\it \bf b}$~\footnote[2]{Email: 
dblack@jlab.org}},

\author{Amir H. Fariborz$^{\it \bf c}$~\footnote[3]{Email: 
fariboa@sunyit.edu}},

\author{Joseph Schechter$^{\it \bf a}$~\footnote[4]{Email:
schechte@physics.syr.edu}}

\affiliation{$^{\bf \it a}$ Department of Physics, 
 Syracuse University, Syracuse, NY 13244-1130, USA,} 

\affiliation{$^{\bf \it b}$ Jefferson Lab,
12000 Jefferson Ave., Newport News,
 VA 23606, USA,} 

\affiliation{$^{\bf \it c}$ Department of Mathematics/Science,
State University of New York Institute of Technology, Utica,
NY 13504-3050, USA.} 

\date{\today}

\begin{abstract}

We study the role of a possible nonet of light scalar mesons 
in the still interesting $\eta \rightarrow 3\pi$ decay process, 
with the primary motivation of learning more about the scalars 
themselves. The framework is a conventional non-linear chiral 
Lagrangian of pseudoscalars and vectors, extended to include 
the scalars. The parameters involving the scalars were previously 
obtained to fit the s-wave $\pi\pi$ and $\pi$K scatterings in the
region up to about 1 GeV as well as the strong decay 
$\eta' \rightarrow \eta \pi\pi$. At first, one might expect a 
large enhancement from diagrams including a light $\sigma(560)$.
However there is an amusing cancellation mechanism which prevents
this from occurring. In the simplest model there is an 
enhancement of about 13 per cent in the \e3p decay rate due to 
the scalars. In a more complicated model which includes 
derivative type symmetry breakers, the cancellation is modified
and the scalars contribute about 30 percent of the total 
decay rate (although the total is not significantly changed). 
The vectors do not contribute much. Our model produces a 
reasonable estimate for the related $a_0(980)-f_0(980)$ mixing 
strength, which has been a topic of current debate. Promising 
directions for future work along the present line are suggested.

\end{abstract}


\maketitle

\section{Introduction}

There has been a revival of interest recently \cite{kyotoconf} 
in the possible existence of a broad scalar meson (sigma) with
a mass in the 560 MeV region and its corresponding nonet 
partners. A large number of workers 
\cite{vanBev}-\cite{Deirdreetal} have found evidence for the 
sigma in models of $\pi \pi$ scattering even though 
it is partially obscured by background. Generally this state is 
considered to be of exotic nature (more complicated than $q 
\bar q$) and hence an important clue to an understanding of 
QCD in its low energy non-perturbative regime. Similarly, 
analyses of $\pi\pi$ , $\pi$K and $\pi\eta$ scattering 
have provided evidence for the existence of the remaining 
members of a possible light scalar nonet: the $\kappa$, 
the $a_0(980)$ and the $f_0(980)$. In fact, the latter 
two states have been well established experimentally 
for some time. Of course, the treatment of such strongly 
interacting processes is inevitably model dependent and 
there are a number of different opinions as to the correct 
approach \cite{kyotoconf}. Thus it is of great interest to 
see whether treatments of the role of scalars in other 
processes using the same models employed in the scattering 
processes above give consistency with experiment.
     
From this point of view we will study the role of possible light 
scalars in the interesting \e3p decay. Typically this process 
has been treated by chiral perturbation theory \cite{Gasser85}, 
in which the possible effects of scalars have been amalgamated 
into effective contact interactions among the pseudoscalars. 
This is probably the most effective way to study the \e3p decay.
However, our goal here is to learn more about the scalars so it 
is natural to keep them rather than integrating them out. 
Also there is a possibility that a light scalar [like the 
$\sigma(560)$] might give an enhancement due to closeness of 
its propagator to the pole [see for instance Feynman diagrams 
like (a) and (b) of Fig.\ref{scdiagrams}]. Another reason for 
including light scalars explicitly is to become more familiar 
with the isospin violating $a_0(980)-f_0(980)$ transition 
which should play a role in the \e3p decay and has also 
recently been postulated \cite{Close01} to provide an 
explanation for observations of anomalously strong 
$a_0(980)$ central production and the large
$\Gamma(\phi \rightarrow f_0 \gamma) 
/ \Gamma(\phi \rightarrow a_0 \gamma)$ ratio. It is important 
to know whether the value consistent with the eta decay 
determination is consistent with these proposed new effects. 
Doubts about whether an unreasonably large value was assumed in 
\cite{Close01} were expressed in \cite{Achasov02}. These doubts 
were confirmed \cite{Black02} using the work of the present 
paper. Still another reason for the interest in the effects 
of the scalars in \e3p is to provide an orientation for the 
discussion of the apparently puzzling $\eta' \rightarrow 3\pi$ 
decays in which light scalar mesons can be reasonably expected 
to have very large effects. We will give only a preliminary 
discussion of this process here.

In section II we give a brief historical outline of treatments 
of \e3p decay based on chiral symmetry. A number of well known 
ambiguities in the analysis are briefly described.

Our calculation is based on the tree level treatment of a chiral 
Lagrangian containing pseudoscalars, vectors and a postulated 
nonet of light scalars. Since the calculation is somewhat 
complicated, it seems to us helpful to present the results in 
a series of steps. First, in section III we give the results of 
using a Lagrangian containing only pseudoscalars with minimal 
symmetry breaking terms.

To this Lagrangian we add, in section IV, the scalar mesons. It 
will be seen that the individual scalar diagrams are quite large 
but there is a lot of cancellation so that the net effect is not 
at all dominant. However the scalars do, as desired, increase 
the predicted decay rate in a noticeable way. Next, the effect 
of adding some derivative type symmetry breakers for the 
pseudoscalars is described in section V. This doesn't much 
change the overall rate but modifies the somewhat delicate 
cancellations so that the scalars end up making a larger 
percentage contribution than before. In low energy calculations 
of this sort one always may expect some contributions from the 
vector mesons. This is discussed in section VI where it is shown 
that, although there is a new type of diagram the vectors do not 
produce a big change in the previous results.

Section VII contains a discussion of the results and directions 
for further work. For the convenience of readers, material 
describing the chiral Lagrangian used is brought together in 
Appendix A. Similarly the detailed expression
for the decay amplitude is given in Appendix B.

\section{Historical Background on the \e3p decay}

The study of \e3p has turned out to be surprisingly 
complicated and correspondingly important for 
understanding the non-perturbative (low energy) 
structure of QCD. Chiral dynamics 
in various forms has been the basic tool.  
Since the process violates G-parity it was 
initially assumed to be of electromagnetic nature, 
mediated by an effective photon exchange operator 
proportional to the product of two electromagnetic 
currents. The old ``current algebra'' approach 
had previously predicted the 
$K_L \rightarrow \pi^+ \pi^- \pi^0$ 
spectrum shape \cite{haranambu} to be 
\begin{equation}
1 - \frac{2E_0}{m},
\label{shape}
\end{equation}
where $m$ is the $K_L$ mass and $E_0$ the energy 
of the $\pi ^0$ in the $K_L$ rest frame.  
This shape, which is in reasonable 
agreement with experiment, resulted from the vanishing 
commutator of the axial charge transforming like a $\pi^+$  
with the appropriate product of two weak currents.  
When Sutherland \cite{Sutherland66} repeated this type 
of calculation for $\eta \rightarrow \pi^+ \pi^- \pi^0$ 
with the product of two electromagnetic currents he found 
that the decay amplitude was actually 
zero (to this leading order).  Thus the \e3p decay did 
not seem to be mediated by a virtual photon emission and
reabsorption. In fact, it was found \cite{bose} that a quark 
scalar density operator with the $\Delta I = 1$ property 
proportional to
\begin{equation}
\bar u u - \bar d d 
\label{scalardensity}
\end{equation}
would give a non-zero result for the decay rate.  A more 
detailed treatment \cite{Chiu67} showed that the 
quark density operator gave the same spectrum for 
$\eta \rightarrow \pi^+ \pi^- \pi^0$  
as in Eq.(\ref{shape}) with $m$ the $\eta$ mass in this case.  
Such a result is in fairly good agreement with experiment.  
The scalar density interaction in Eq.(\ref{scalardensity}) 
was recognized \cite{su} to be the fundamental up-down quark 
mass difference generated by the Higgs boson in 
the electroweak theory.  

However, the predicted rates of the 
$\eta \rightarrow \pi^+ \pi^-\pi^0$  and 
$\eta \rightarrow 3 \pi^0$ modes (both the ratios 
and the absolute values) did not agree well with experiment at 
that time. Some years later, after more precise experiments, 
the ratio of the rates for $\pi^+ \pi^- \pi^0$ to $3\pi^0$ 
modes stabilized around the value expected from isospin 
invariance. On the other hand the absolute rate has only 
recently stabilized to a value notably larger than that 
predicted by theory. The theory behind the current algebra 
results could be economically presented in the framework of an 
effective chiral Lagrangian. For most low energy processes 
where the scheme could be expected to work, the tree level 
computation did produce results within 25 $\%$ or so of 
experiment. Thus the relatively poor prediction for
\e3p at tree level is somewhat surprising.

An improvement was obtained by Gasser and Leutwyler 
\cite{Gasser85} who carried the computation of the chiral 
Lagrangian amplitude to one loop level. Since the non-linear 
chiral Lagrangian is non-renormalizable, this required the 
addition of new counterterms. Their finite parts were
new parameters which could be mostly determined from other 
processes. They obtained the result $\Gamma(\eta 
\rightarrow \pi^+ \pi^- \pi^0) = 160 \pm 50$ eV which 
may be compared with the present experimental value \cite{pdg}  
${\Gamma \left( \eta \rightarrow \pi^+ \pi^- 
\pi^0 \right)}_{\rm expt} = 267 \pm 25$ eV. The extra effects
included involve both the implicitly integrated-out heavier 
meson exchanges and partial unitarization to one loop order.  
One might expect a two loop calculation in the chiral 
perturbation scheme to be valuable but this may involve too 
many unknown parameters at the present stage. A dispersion 
approach using the Gasser-Leutwyler result as a subtraction
gave an improved estimate \cite{Kambor96} $\Gamma 
\left( \eta \rightarrow \pi^+ \pi^- \pi^0 \right) = 209 \pm 20$ 
eV, which still seems too small.   

A possible source of ambiguity arises from the determination of 
the coefficient of the driving scalar density interaction in
Eq.(\ref{scalardensity}). This is determined from the 
$K^0 - K^+$ mass difference, which in turn has two components

\begin{eqnarray}
m^2 (K^0) - m^2 (K^+) &=& 
{ \left[ m^2 (K^0) - m^2 (K^+) \right] }_{\rm
quark \, mass}
+ { \left[ m^2 (K^0) - m^2 (K^+) \right] }_{\gamma},
\label{deltak}
\end{eqnarray}
corresponding to the quark mass differences and the virtual photon
emission and reabsorption diagrams respectively. The latter is 
given in the chiral limit by $m^2 (\pi^0) - m^2 (\pi^+)$ according 
to Dashen's theorem \cite{dashen} and the reasonable assumption 
that the photon contribution saturates the pion mass difference.  
A number of authors \cite{Donoghue92} have argued that there are 
important corrections to Dashen's theorem which have the effect of 
boosting the \e3p decay rate.  

If one questions Dashen's theorem it is natural to also question
Sutherland's result, which deals with the direct electromagnetic
contribution to \e3p . An investigation of this point yielded
\cite{Baur96} the estimate that there was only about a 2$\%$ 
change arising from this, although it decreased rather than raised 
the rate. 

Still another point which may repay further investigation concerns 
the possible subtleties arising from $\eta - \eta^\prime$ mixing.  
An understanding of the $\eta^\prime \rightarrow 3 \pi$ process, 
for example, might clarify this point. This process has been 
treated by some authors \cite{hudnall}, \cite{paula} in the 
literature but has received only a fraction of the attention 
given to \e3p .

In the present paper we will focus on learning more about the 
putative nonet of light scalar mesons by studying their 
contribution to \e3p. 

\section{Chiral symmetry results to lowest order}

For comparison, we first present the well-known results when 
only the terms present in the lowest order chiral Lagrangian of 
pseudoscalars are kept.

\begin{eqnarray}
{\cal L}_{LO} &=&
\frac { {F_\pi}^2}{8} {\rm Tr} \left( \partial_\mu U
\partial _\mu U^\dagger \right) 
+ \delta^\prime {\rm Tr} \left[ {\cal M} 
\left( U + U^\dagger \right) \right] 
+ \frac{\kappa}{576} {\rm
ln}^2 \left( \frac { {\rm det} U}{{\rm det} U^\dagger} \right),
\label{lowestorderLag}
\end{eqnarray}
where the last term [see Eq.(\ref{extraetaprime}) and comments 
there] supplies mass to the SU(3) singlet state and $\cal M$ is 
defined in (\ref{spurion}). Fitting ${\cal L}_{LO}$ to the 
experimental masses determines $\delta^\prime = F_\pi^2m_\pi^2/8$.

\begin{figure}[htbp]
\begin{center}
\epsfxsize = 8.5cm
\ \epsfbox{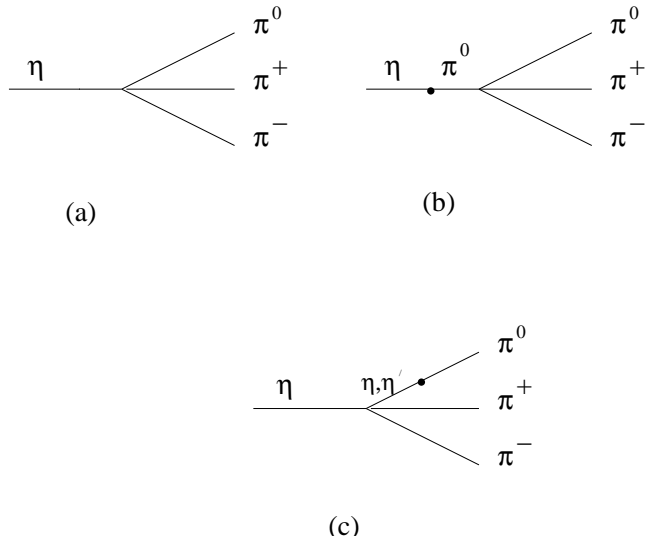}
\end{center}
\caption
{Feynman diagrams representing the pseudoscalar meson contribution
to the decay $\eta \rightarrow \pi^+ \pi^- \pi^0$ .}
\label{psdiagrams}
\end{figure}

The $\eta \rightarrow \pi^+ \pi^- \pi^0 $ amplitude receives, in 
this approximation, contributions from diagrams (a), (b) and (c) 
of Fig.\ref{psdiagrams}, which are given in Eq.(\ref{psampls}) 
(with the non leading corrections deleted). To a reasonable 
approximation which displays the key dependences these sum up to 
the lowest order result for the  
$\eta \rightarrow \pi^+\pi^- \pi^0 $ amplitude
\begin{equation}
M_{0+-}(E_1,E_2,E_3) 
\approx \frac{16 i \delta^\prime y}{F_\pi^4} 
{\rm cos} \theta_p (1 - \frac {2E_1}{m_\eta} ) .
\label{lowestorderAmp}
\end{equation}
Here $E_1$ is the $\pi^0$ energy in the $\eta$ restframe
and $y$ is the dimensionless parameter in Eq.(\ref{spurion})
which measures the isospin violation in the quark mass matrix.
Assuming Dashen's theorem, Eq.(\ref{lowestorderLag}) yields
\begin{equation}
8\delta^\prime y =
F_\pi^2(m_{K^0}^2-m_{K^+}^2-m_{\pi^0}^2+m_{\pi^+}^2) ,
\end{equation}
which allows us to solve for $y$. Furthermore $\theta_p$ is 
the ``nonstrange-strange" pseudoscalar mixing angle defined in 
Eq.(\ref{eta_etap}); it is generally taken to be about $37^o$. 
It is related to the ``octet-singlet" angle, $\theta$ by
\begin{equation}
{\rm cos}\theta_p = \frac{{\rm cos}\theta 
-\sqrt 2{\rm sin}\theta}{\sqrt 3} .
\end{equation}
Then Eq.(\ref{lowestorderAmp}) agrees with Eq. (1.14) of 
\cite{Gasser85} except that they neglected $\eta-\eta^\prime$ 
mixing by replacing ${\rm cos}\theta_p \rightarrow 1/\sqrt 3$ 
in what was denoted the current algebra formula. The matrix 
element for $\eta \rightarrow 3\pi^0$ is given in general by 
\begin{equation}
M_{000} = M_{0+-}(E_1,E_2,E_3) + M_{0+-}(E_2,E_1,E_3) +
M_{0+-}(E_3,E_2,E_1).
\end{equation}
The widths are then, defining $\Gamma_{0+-}=
\Gamma(\eta \rightarrow \pi^+ \pi^- \pi^0)$ and 
$\Gamma_{000}=\Gamma (\eta \rightarrow 3\pi^0)$:
\begin{eqnarray}
\Gamma_{0+-}&=&
\frac{1}{64 \pi^3 m_\eta} \int {\rm d}E_1 {\rm d}E_2
{\left| M_{0+-} \right|}^2, \nonumber \\
\Gamma_{000}&=&
\frac{1}{384 \pi^3 m_\eta} \int {\rm d}E_1 {\rm d}E_2
{\left| M_{000} \right|}^2.
\end{eqnarray}

Using ${\cal L}_{LO}$, with parameters determined as described 
above, we get the tree-level results [from the first terms in 
each of $M_{contact}^{a,b,c}$ in Eqs.(\ref{psampls})]:
\begin{eqnarray}
\Gamma_{0+-} &=& 106 \, {\rm eV} ,\nonumber \\
\frac{\Gamma_{000}}{\Gamma_{0+-}} &=& 1.40 .
\label{eta_LO_result}
\end{eqnarray}

These may be compared with the experimental results \cite{pdg}
\begin{eqnarray}
{(\Gamma_{0+-})}_{\rm expt} &=& 267 \pm 25 
\hskip.2cm {\rm eV} ,\nonumber \\
{(\frac{\Gamma_{000}}{\Gamma_{0+-}})}_{\rm expt} &=& 
1.40 \pm 0.01 ,
\label{eta_experimental_rate}
\end{eqnarray}
which demonstrate the disagreement with experiment for the
overall rates in the simplest model. However the width ratio 
has about the correct magnitude. The related energy spectrum is 
also about the correct magnitude. The squared matrix element 
is usually described by quantities 
$a$, $b$ and $c$ defined from 
\begin{equation} 
{\left| M_{0+-} \right|}^2 \propto 
( 1 + aY + bY^2 + cX^2 \ldots ) ,
\label{spectrum}
\end{equation}
with $X = \frac{\sqrt{3}}{m_\eta - 3m_\pi} (E_2 - E_3)$ and 
$Y=\frac{3}{m_\eta - 3 m_\pi} ( E_1 - m_\pi) - 1$. 
In the present paper we shall not take into account the 
(not completely negligible) kinematic $\pi^0-\pi^+$ mass 
difference. See \cite{Gasser85} for a discussion of this point.  
The predictions from this simple model, $a\approx-1$ and 
$b\approx0.25$ are similar to the experimental results 
\cite{Abele} $a_{\rm exp} = -1.19
\pm 0.07$ and $b_{\rm exp} = 0.19 \pm 0.11$ with $c$=0.  

It is of some interest to also give the predictions for the
$\eta^\prime \rightarrow 3 \pi$ decay process at tree level using 
the simple Lagrangian Eq.(\ref{lowestorderLag}). It just is 
necessary (see Appendix B) to replace ${\rm cos} \theta_p$ by 
${\rm sin} \theta_p$ and $m_\eta$ by $m_{\eta^\prime}$ in 
Eq.(\ref{lowestorderAmp}) to get for the $\eta' \rightarrow 
\pi^0 \pi^+ \pi^-$ matrix element: 
\begin{equation}
M_{0+-}^\prime(E_1,E_2,E_3) 
\approx \frac{16 i \delta^\prime y}{F_\pi^4} {\rm
sin} \theta_p (1 - \frac {2E_1}{m_{\eta^\prime}} ).
\end{equation}
This leads to the predictions for the $\eta^\prime$ modes:
\begin{eqnarray}
\Gamma^\prime_{0+-} &= 497 \, {\rm eV} ,  \nonumber \\
\Gamma^\prime_{000} &= 562 \, {\rm eV} .
\label{etap_LO_result}
\end{eqnarray}
The experimental results are given as \cite{pdg} 
\begin{eqnarray}
{(\Gamma^\prime_{0+-})_{\rm exp}} & 
< {10}^4 \, {\rm eV} , \nonumber \\
{(\Gamma^\prime_{000})}_{\rm exp} & = 315  \pm 56 \, {\rm eV}.
\label{etap_experimental_rate}
\end{eqnarray}
Only the $3\pi^0$ mode has really been measured; its width is 
smaller than predicted in the simple model just 
presented. One would, of
course, expect better agreement for the low energy process
$\eta \rightarrow 3 \pi$ for which chiral perturbation theory 
should be more clearly reliable. In the present paper we shall 
just make a few remarks on this more complicated process.  

It may also be worthwhile to give a rough estimate of the 
corrections to the rates corresponding to violations of Dashen's 
Theorem mentioned earlier. If we parameterize the electromagnetic 
contribution to the $K^+ - K^0$ mass difference as 
\begin{equation}
{(m_{K^0}^2 - m_{K^+}^2 )}_\gamma = f (m_{\pi^0}^2 - m_{\pi^+}^2 ),
\label{definef}
\end{equation}
where $f = 1$ corresponds to Dashen's Theorem, we would find by
using Eq.(\ref{deltak}) that the $\eta, \eta^\prime 
\rightarrow 3 \pi$ rates predicted for ${\cal L}_{LO}$ should be 
multiplied by 
\begin{equation}
{ \left[ \frac { (m_{K^0}^2 - m_{K^+}^2 ) 
- f (m_{\pi^0}^2 - m_{\pi^+}^2 )} 
  { (m_{K^0}^2 - m_{K^+}^2 ) 
- (m_{\pi^0}^2 - m_{\pi^+}^2) } \right]
  }^2 .
\label{Dashen_violation_factor}
\end{equation}
For $f \approx 2$, which was actually found many years ago 
\cite{Socolow65} the correction factor is about $1.54$ and would 
give $\Gamma_{0+-} \approx 163$ eV. This corresponds to the
overall factor, $y$ taking the value $-0.33$ while the Dashen's 
theorem value used to obtain Eqs.(\ref{eta_LO_result}) and 
(\ref{etap_LO_result}) is $y=-0.277$.

\begin{figure}[htbp]
\begin{center}
\epsfxsize =8.5cm
\ \epsfbox{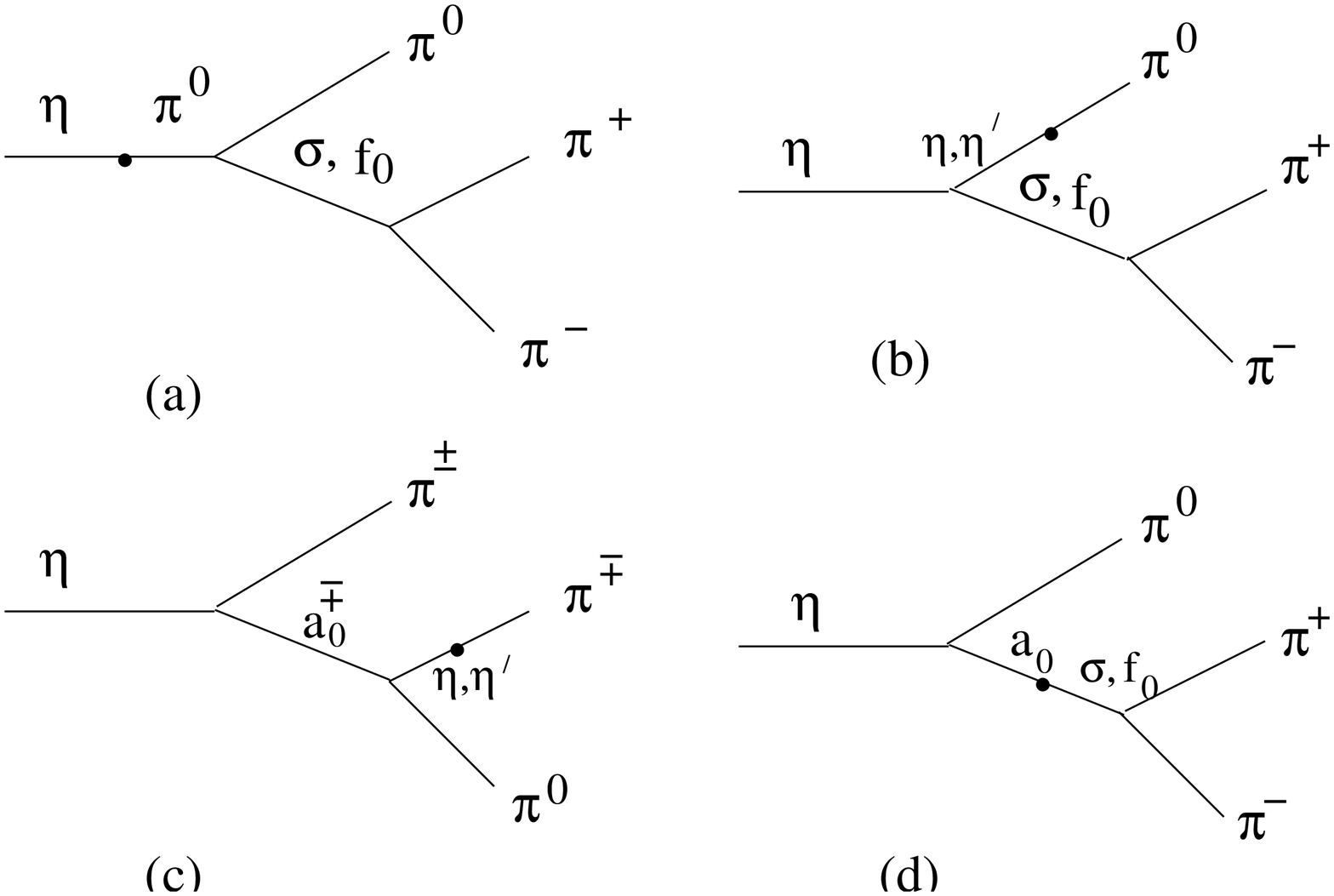}
\end{center}
\caption
{Feynman diagrams representing the scalar meson contributions
to the decay $\eta \rightarrow \pi^+ \pi^- \pi^0$ .}
\label{scdiagrams}
\end{figure}

\begin{figure}[htbp]
\begin{center}
\epsfxsize = 8.5cm
\ \epsfbox{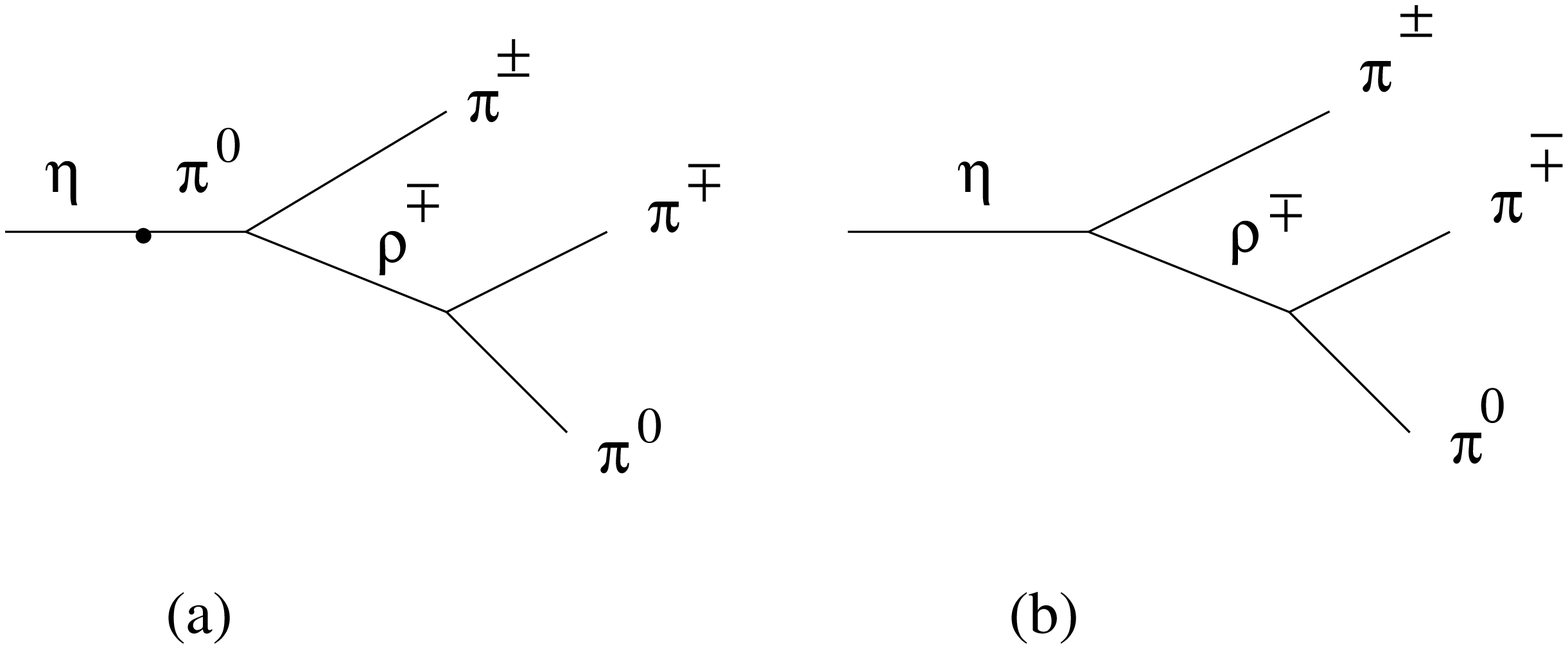}
\end{center}
\caption
{Feynman diagrams representing the $\rho$ meson contributions
to the decay $\eta \rightarrow \pi^+ \pi^- \pi^0$ .}
\label{vmdiagrams}
\end{figure}

\section{Including light scalar meson interactions}

Now we will study what effects the inclusion of a nonet of light
scalar mesons will have on the \e3p calculation. We designate the
scalar nonet by the $3 \times 3$ matrix $N_a^b$ whose interactions
are also listed in Appendix A.\quad $N$ is assumed to contain the
well-established $f_0(980)$ isoscalar and the $a_0(980)$ isovector as
well as the $\sigma(560)$ and the strange $\kappa(900)$. 
Of these only the $\kappa(900)$ will not contribute 
to \e3p at tree level. The quark structure of such a 
nonet has been the subject of much
discussion \cite{kyotoconf}-\cite{Deirdreetal}.
If this were an ideal nonet like $\rho - \omega - K^*- \phi$ 
one would expect the roughly degenerate $a_0(980)$ and
$f_0(980)$ to be lowest rather than highest in mass. Actually 
the masses are better understood intuitively \cite{jaffe} if 
$N_a^b$ is an ``ideal dual nonet'' constructed 
as $Q_a {\bar Q}^b$ with 
$Q_a \sim \epsilon_{abc} {\bar q}^b {\bar q}^c$;  $q_a$ being 
the ordinary quark. Then the observed inverted mass ordering 
is easily seen to follow just from counting the number of strange 
quarks in $N^a_b$. It is important to note that the form of the 
couplings of $N_a^b$ to the particles of the non-linear chiral 
Lagrangian being used depend only on the flavor transformation 
properties of $N_a^b$. This does not distinguish different quark 
substructures. What is sensitive to the quark substructure is the 
scalar mixing angle, $\theta_s$, defined from 
\begin{equation}
\left( \begin{array}{c} \sigma\\ f_0 \end{array} \right) = \left(
\begin{array}{c c} {\rm cos} \theta_s & 
-{\rm sin} \theta_s \\ {\rm sin}
\theta_s & {\rm cos} \theta_s \end{array} \right) 
\left( \begin{array}{c}
N_3^3 \\ \frac {N_1^1 + N_2^2}{\sqrt 2} \end{array} \right).
\label{scalar-mixing-convention}
\end{equation}
Small values of $\theta_s$ would typify a dual ideal nonet while 
$\left| \theta_s \right|$ about $\frac{\pi}{2}$ would typify a
conventional nonet. Fitting the $\pi \pi$ and $\pi$K scattering
amplitudes, including the effects of these scalar resonances, selects
\cite{BFSS2} the small value $\theta_s = -20.3^o$.

The scalar nonet mass terms in Appendix A are specified by the four
parameters $(a,b,c,d)$. The needed chiral invariant scalar
-pseudoscalar-pseudoscalar $S \phi \phi$-type couplings are 
specified by the parameters $(A,B,C,D)$ and the mixing
angle $\theta_s$. These were all determined from fitting to 
$\pi \pi$ scattering, $\pi$K scattering and the strong decay 
$\eta^\prime \rightarrow \eta \pi \pi$. 

Actually there is reason to believe \cite{2nonetmixing} 
that the scalars may be best understood as mixtures of 
a lighter dual nonet and a heavier ordinary nonet. 
From that point of view, which will be explored more 
fully in the future, the present single 
nonet, $N_b^a$ should be regarded as an approximation to the 
situation where the heavier (after mixing) particles have been 
integrated out.  

The Feynman diagrams for the scalar contributions to 
$\eta \rightarrow \pi^+ \pi^- \pi^0$ are shown in 
Fig.\ref{scdiagrams}. Notice that the diagram in (d) involves 
new $a_0 - \sigma$ and $a_0 - f_0$ isospin violating transitions 
rather than the $\pi^0 - \eta$ and $\pi^0 - \eta^\prime$ transitions 
which play an important role in the other diagrams. Their
strengths $A_{a\sigma}$ and $A_{af}$ (see Appendix B) were determined
simply by including the effects of isospin violation contained in the
spurion matrix $\cal M$ in the $b$ and $d$ scalar mass terms.
Therefore, this does not introduce any new parameters. Actually, the
possibility of such contributions was suggested a long time ago
\cite{hudnall} as a possible solution of the \e3p width problem.
Recently a relatively large $a_0 - f_0$ mixing has been 
suggested \cite{Close01} as a way of understanding both anomalously 
large $a_0$ central production and the large 
$\Gamma(\phi \rightarrow f_0 \gamma) 
/ \Gamma (\phi \rightarrow a_0 \gamma)$ ratio.  
However criticisms of this explanation have also been 
presented \cite{Achasov02}. Clearly it may be useful 
for studies of processes other than 
$\eta \rightarrow 3\pi$ to give the coefficients
of the scalar isospin violating two point Lagrangian,
\begin{equation}
{\cal L} = A_{a\sigma}a_0^0 \sigma + A_{af}a_0^0f_0,
\label{scalar_isospin_violating_transitions}
\end{equation}
determined consistently with the $\eta \rightarrow 3\pi$
calculation. Using the parameters from 
Eq.(\ref{abcd-parameters}) in Eq.(\ref{two-point-vertices})
we find
$A_{a\sigma}=0.0170 y$ ${\rm GeV}^2$, $A_{af}=0.0234 y$ 
${\rm GeV}^2$, where $y$ is the quark mass ratio 
$\frac{m_u-m_d}{m_u+m_d}$. Notice that $y$ (which is negative) is 
an overall factor for the $\eta \rightarrow 3\pi$ amplitude in 
the present model.

\begin{figure}[htbp]
\begin{center}
\epsfxsize = 6.5cm
\ \epsfbox{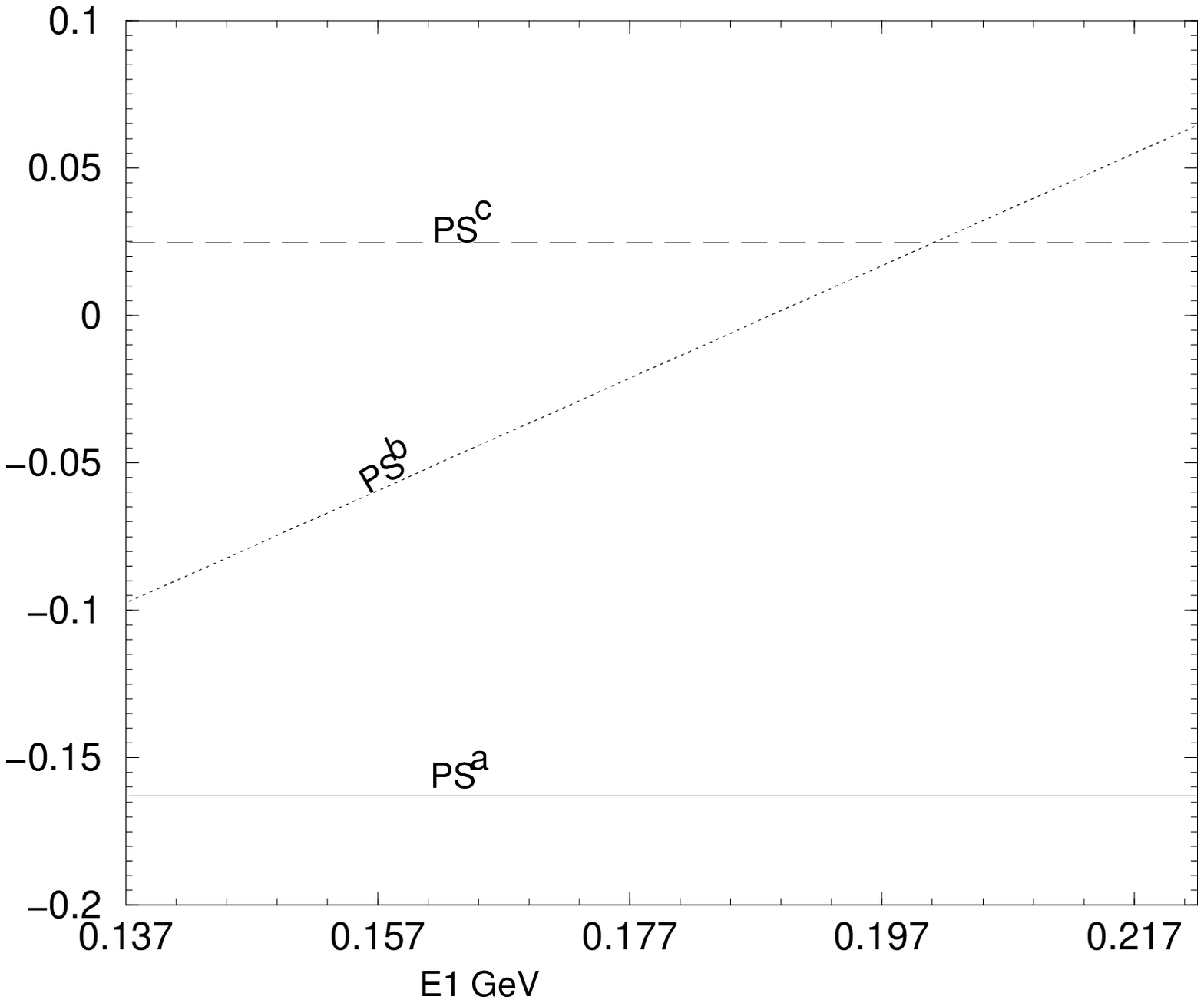}
\epsfxsize = 6.5cm
\ \epsfbox{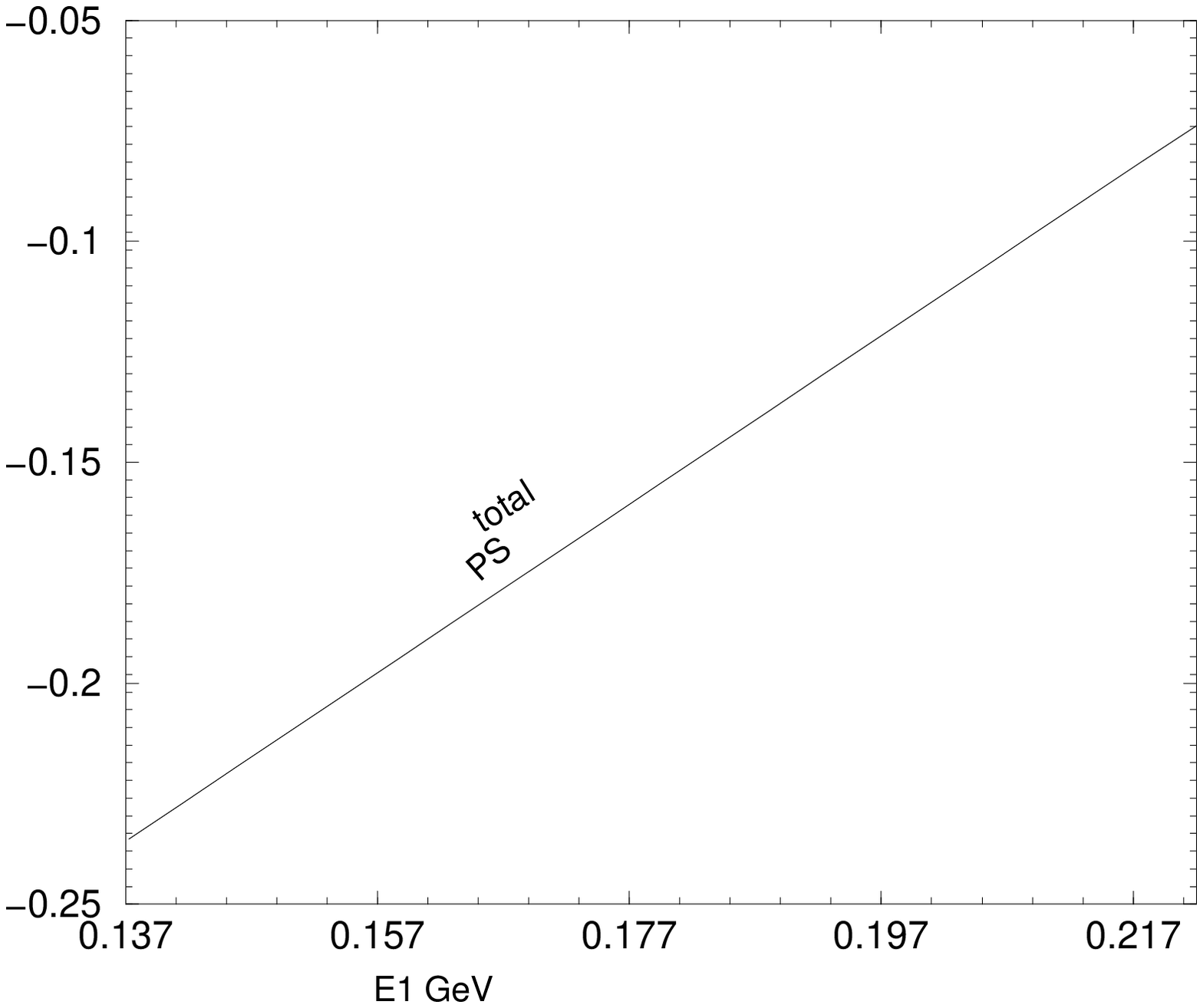}
\end{center}
\caption
{Plot of different contributions to the leading order
$\eta \rightarrow \pi^+ \pi^- \pi^0$ amplitude
as a function of the energy of the neutral pion.
On the left, the solid line corresponds to
the direct four-point isospin-violating
$\eta \pi^+ \pi^- \pi^0$ vertex of Fig.\ref{psdiagrams}a.
The energy-dependent dotted line is due to an
$\eta \pi^0$ transition followed by a four-pion contact
vertex (Fig.\ref{psdiagrams}b).  The dashed line is the small
contribution due to Fig.\ref{psdiagrams}c. (with both $\eta$ and
$\eta^\prime$ included in the final line).
On the right we show the total leading order amplitude,
which is the sum of Figs.\ref{psdiagrams}a--\ref{psdiagrams}c.}
\label{pseudoscalar01fig}
\end{figure}

We would like to discuss the effects of the scalars when added
to the more realistic Lagrangian presented in Appendix A
which contains both pseudoscalars and vectors. This Lagrangian 
contains additional symmetry breaking terms 
($\alpha_p$ and $\lambda^{\prime}$) to account for the ratio
of pseudoscalar decay constants, $F_K/F_\pi$ being different
from unity as well as a number of terms describing the properties
of the vector mesons. Of course, the vector mesons play an important
role in low energy processes. Since there are many terms it seems 
useful to add these new features one at a time. Thus in the present 
section we will consider just the minimal pseudoscalar Lagrangian, 
${\cal L}_{LO}$ [Eq.(\ref{lowestorderLag})], to be present in 
addition to the scalars. Furthermore, it is instructive to look at 
the contributions to the amplitude from different diagrams in 
order to see how they combine to give the predicted total  
$\eta \rightarrow \pi^+ \pi^- \pi^0$ width. In section III we 
reviewed the leading order calculation of the \e3p amplitude 
which gives the result 
$\Gamma (\eta \rightarrow \pi^+ \pi^- \pi^0) = 106$ eV in the Dashen's
theorem limit. In Fig.\ref{pseudoscalar01fig} we show how the 
individual contributions of the diagrams in Fig.\ref{psdiagrams}
combine to give the leading order amplitude. The magnitude of the
$\eta - \pi^0$ and $\eta^\prime - \pi^0$ transition coefficiencts
[Eq.(\ref{two-point-vertices}) 
with $\alpha_p = \lambda^\prime = 0$, so independent 
of which state is on-shell] are (in ${\rm GeV}^2$):
\begin{equation}
C_{\pi \eta} \approx 0.0042, \quad \quad C_{\pi \eta^\prime} \approx 
0.0031.
\label{etapitrans}
\end{equation}

\begin{figure}[htbp]
\begin{center}
\epsfxsize = 6.5cm
\ \epsfbox{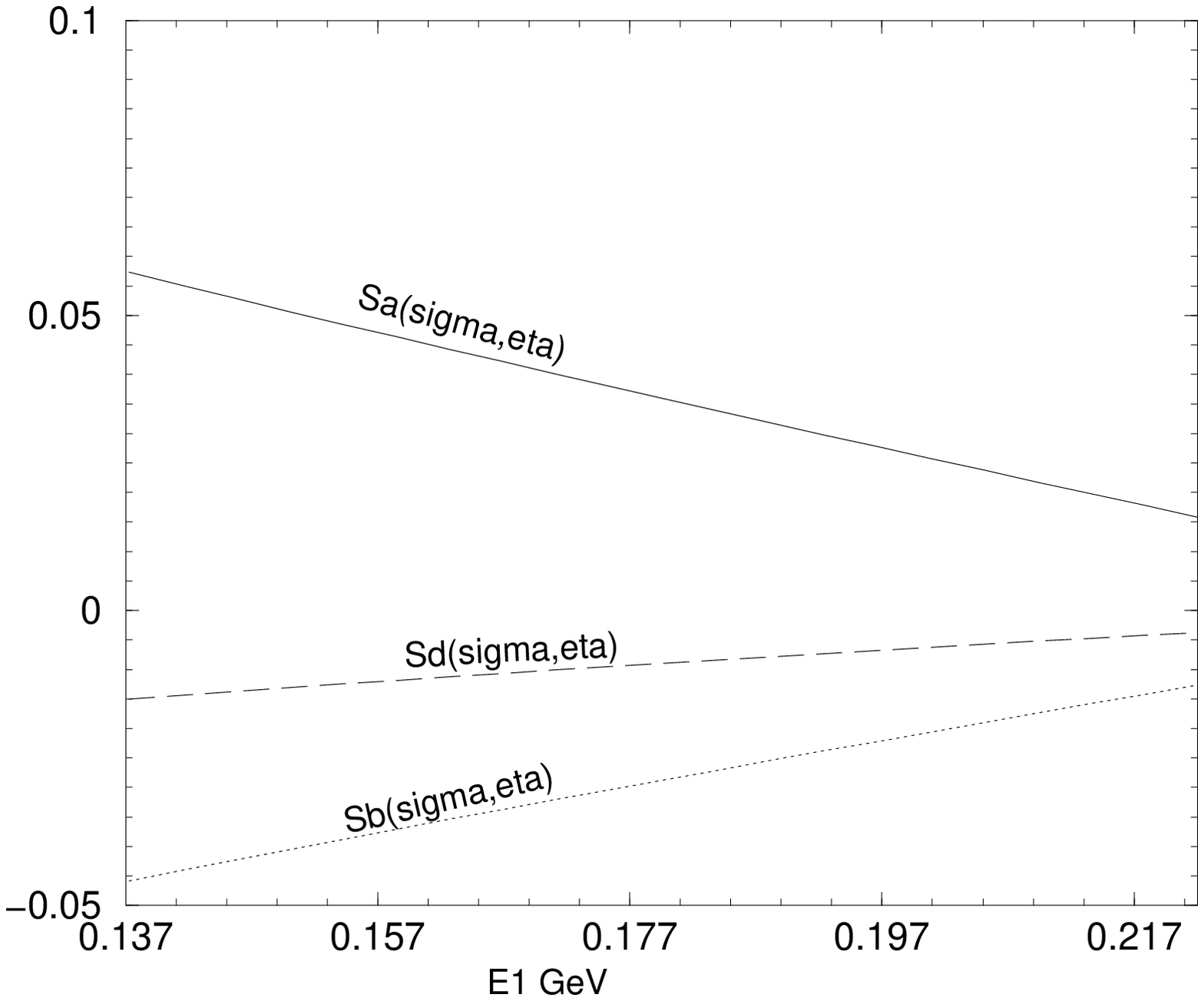}
\epsfxsize = 6.5cm
\ \epsfbox{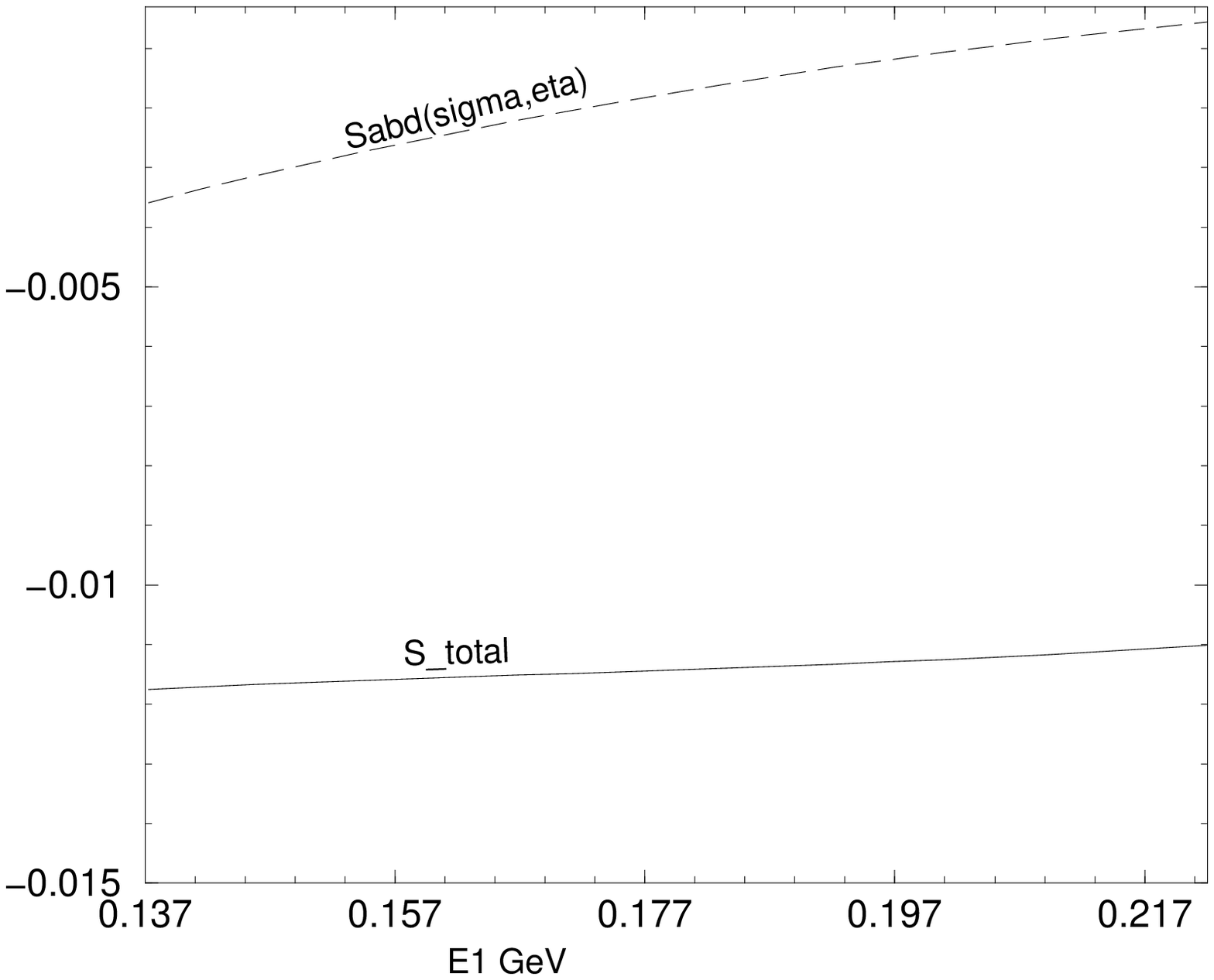}
\end{center}
\caption
{Scalar meson contributions to the
$\eta \rightarrow \pi^+ \pi^- \pi^0$ amplitude
as a function of the energy of the $\pi^0$. On the
left we show the individually 
largest contributions -- the solid line
corresponds to Fig.\ref{scdiagrams}a with 
$\sigma$ exchange, the dotted
line corresponds to Fig.\ref{scdiagrams}b 
with $\sigma$ exchange and the $\eta$-$\pi^0$ 
transition while the dashed line corresponds 
to Fig.\ref{scdiagrams}d involving the new
isospin violating $a_0 - \sigma$ transition.   
On the right we show the total amplitude due 
to the scalar mesons alone. The solid line 
is due to all of the diagrams in Fig.\ref{scdiagrams}
(for the sample value $E_2 = m_\pi$)
and the dashed line is the sum of the three
largest amplitudes plotted on the left 
and discussed in the text.}
\label{scalar_pseudoscalar01fig}
\end{figure}

Three of the twelve scalar diagrams in Fig.\ref{scdiagrams} can be 
seen to be larger (by at least an order of magnitude) than all the rest.  
These are the three diagrams involving the lightest of the scalar 
mesons, $\sigma$, and are contained in Figs.\ref{scdiagrams}a, 
\ref{scdiagrams}b and \ref{scdiagrams}d. For the case of 
Fig.\ref{scdiagrams}b the graph with an $\eta - \pi^0$
transition dominates that with an $\eta^\prime$ transition
because the latter is suppressed by the (square of) the
$\eta^\prime$ mass in the denominator of the propagator
and also the smallness of the associated coupling 
constants/transition coefficients. 
In Fig.\ref{scalar_pseudoscalar01fig} we present the
$\eta \rightarrow \pi^+ \pi^- \pi^0$ amplitudes arising from the three 
diagrams just mentioned and notice that they cancel almost completely.  
In particular the $\sigma$ exchange diagrams in 
Figs.\ref{scdiagrams}a and \ref{scdiagrams}b 
have opposite signs, as expected -- their structure 
is roughly similar except the propagators have a relative minus sign.  
We note also that the new isospin violating diagram, 
involving an $a_0 - \sigma$
transition, turns out not to lead to dramatically larger contributions
than the other diagrams. 
The cancellation between different diagrams involving 
the sigma means that the total scalar contribution to the 
$\eta \rightarrow \pi^+ \pi^- \pi^0 $  
width is smaller than might be expected and 
in fact arises mainly from the 
$a_0^{\pm}$ exchanges in Fig.\ref{scdiagrams}c. 
Comparing Fig.\ref{scalar_pseudoscalar01fig}b with 
Fig.\ref{pseudoscalar01fig}b 
shows that the net scalar contribution does enhance the overall 
$\eta \rightarrow \pi^+ \pi^- \pi^0$ rate. 
Specifically, including the 
scalar contributions with the pseudoscalar Lagrangian 
$\cal L_{LO}$ has increased $\Gamma_{0+-}$ 
by 16 per cent, from 106 eV to 124 eV. The
ratio $\Gamma_{000}/\Gamma_{0+-}$ is essentially unchanged.

Actually, the calculations above have neglected the finite
widths of the $\sigma$, $f_0$ and $a_0$ particles. We take these
into account by making the replacements in the corresponding
propagators [see Eq.(\ref{scampls})]:

\begin{equation}
\frac{1}{m_X^2 +q^2} \rightarrow \frac{1}{m_X^2 + q^2 -im_X\Gamma_X},
\end{equation}
where X stands for $\sigma$, $f_0$ or $a_0$. The $\Gamma_X$ are
given in Appendix A. These replacements modify the result
to $\Gamma_{0+-}= 120$ eV, a 13 per cent increase relative 
to the leading order result. The width effect is mainly due to 
the $\sigma$ propagator.

We may note that the improvement due to the scalars is consistent
with the lower values of the prediction, $160 \pm 50$ eV obtained
from the next order of chiral perturbation theory 
in ref.\cite{Gasser85}. 
The numerical amount of suppression of the scalar contribution 
to the decay rate from cancellation of Figs. (a) and (b) of 
Fig.\ref{scdiagrams} is due to the fitted values of 
$\gamma_{\sigma \pi \pi}$ and $\gamma_{\sigma \eta \eta}$ given in 
Eq.(\ref{scalar_vertices}). If we wanted to raise the predicted 
rate to about 150 eV (still keeping Dashen's theorem in the
evaluation of $y$) it would be necessary to raise 
$\gamma_{\sigma \eta \eta}$ to about 10.

\section{Effects of Higher Order Pseudoscalar Symmetry Breakers}

So far we have worked only with the leading order chiral 
Lagrangian of pseudoscalars and scalars and 
obtained  (to linear order in $y = - \frac{m_d - m_u}{m_d + m_u}$)
isospin-breaking amplitudes proportional
to $\delta^\prime, b$ and $d$ in Eq. (\ref{LagSB})
of Appendix A. In order to better fit the properties of 
the pseudoscalar mesons we consider, as mentioned above, the
higher-order symmetry breaking terms in Eq.(\ref{LagSB}) 
with coefficients $\alpha_p$ and $\lambda^\prime$.
The numerical values of these parameters are obtained 
in section 4 of Appendix A, based on an overall fitting 
of pseudoscalar meson properties.

\begin{figure}[htbp]
\begin{center}
\epsfxsize = 6.5cm
\ \epsfbox{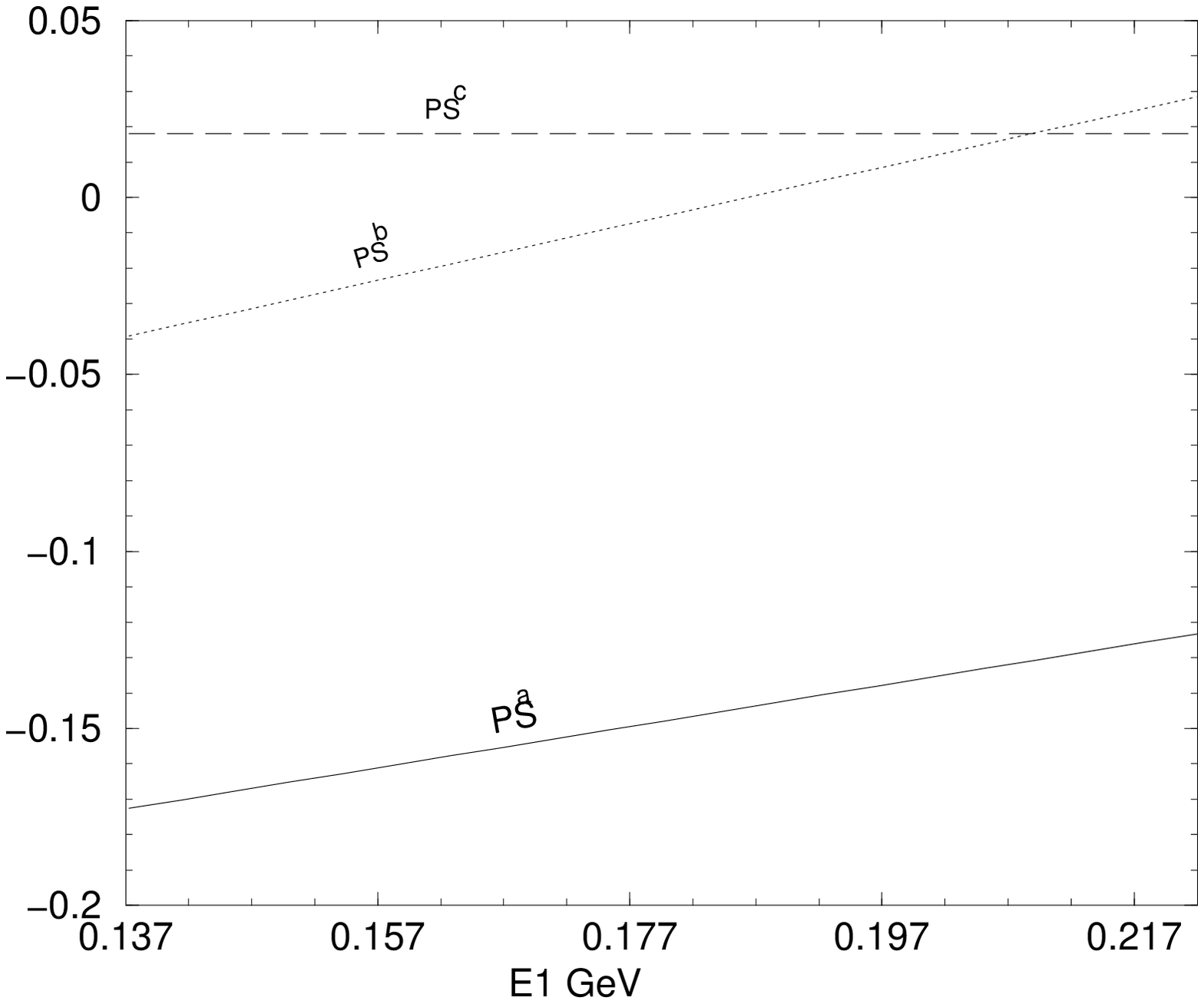}
\epsfxsize = 6.5cm
\ \epsfbox{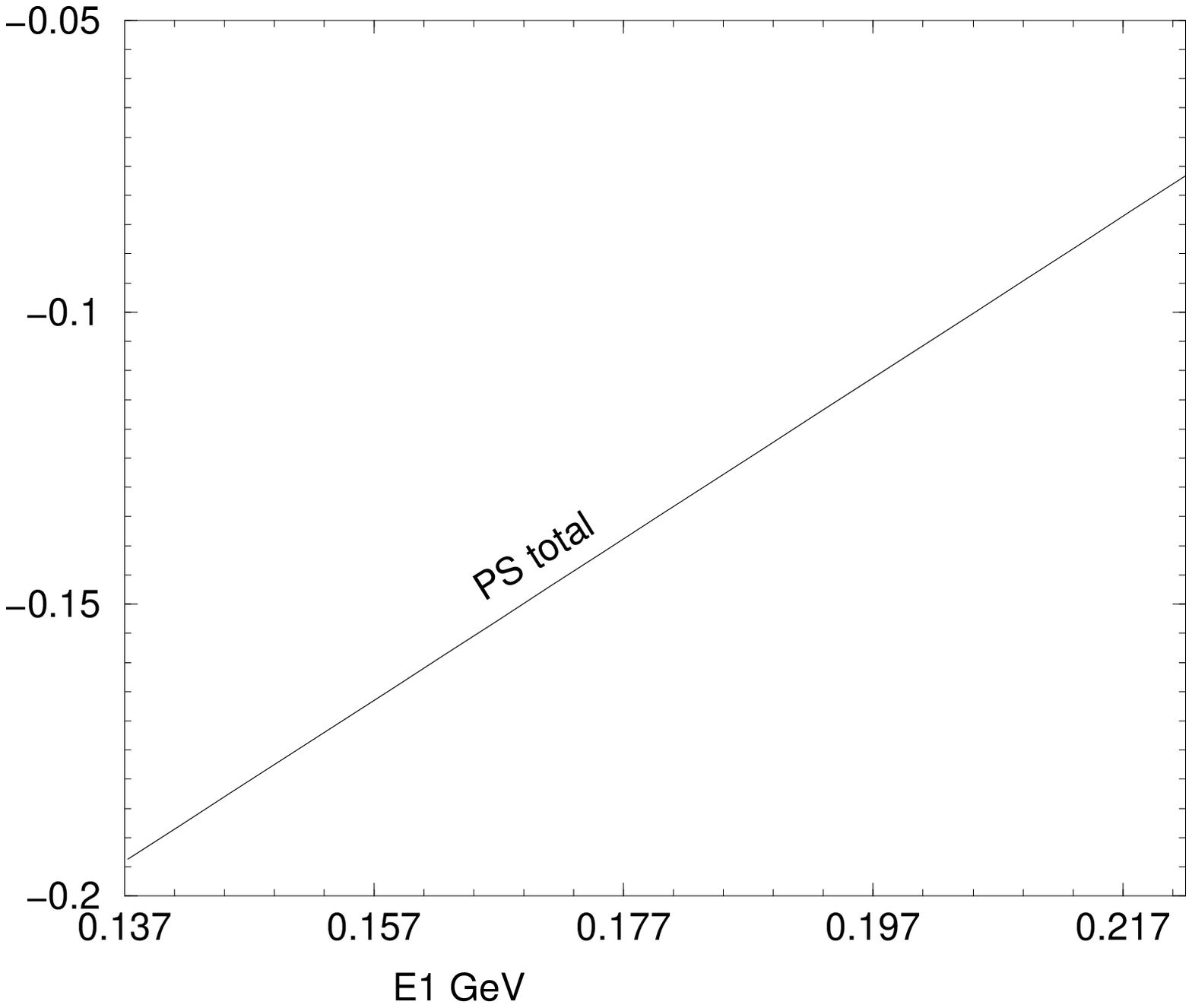}
\end{center}
\caption
{Left:  Contributions due to pseudoscalar mesons alone,
taking into account higher order symmetry breaking effects
encoded in $\alpha_p$ and $\lambda^\prime$.
Compare with the first of Figs.\ref{pseudoscalar01fig}.
Right: Total amplitude due to pseudoscalar mesons alone,
taking into account $\alpha_p$ and $\lambda^\prime$ terms
(compare with second of Figs.\ref{pseudoscalar01fig}).}
\label{scalar_pseudoscalar_alphap01fig}
\end{figure}

Next we examine the effects of these two new symmetry
breaking terms on our previous calculation. It will be 
seen that these effects include an interesting
redistribution of the contributions from the scalar and
pseudoscalar diagrams to the total amplitude. First, the 
contact diagram Fig.\ref{psdiagrams}a will receive corrections
due to the $\alpha_p$ and $\lambda^\prime$
terms [see Eq.(\ref{psampls})]. This results in some 
energy dependence since $\alpha_p$ gives a four-point 
derivative coupling. More importantly, the $\eta - \pi^0$ 
and $\eta^\prime -\pi^0$ transition coefficients relevant 
for \e3p now depend on which particle is on-shell and are 
numerically (in ${\rm GeV}^2$):
\begin{eqnarray}
C_{\pi\eta}^\eta \approx -0.00583y,
 & \quad &  C_{\pi\eta}^\pi \approx -0.0151y \nonumber \\
C_{\pi\eta^\prime}^\pi & \approx & -0.0113y.
\label{newetapi}
\end{eqnarray}
Since $C_{\pi\eta}^\eta$ is now considerably suppressed 
in magnitude, the Feynman amplitude for Fig.\ref{psdiagrams}b 
is now suppressed, while $C_{\pi\eta}^\pi$ and the amplitude 
for Fig.\ref{psdiagrams}c remain about the same. These results, 
due to only pseudoscalars, are summarized in
Fig.\ref{scalar_pseudoscalar_alphap01fig} which may
be compared with Fig.\ref{pseudoscalar01fig}.
The net result is that the total 
$\eta \rightarrow \pi^+ \pi^- \pi^0$ amplitude (shown in 
the second of Fig.\ref{scalar_pseudoscalar_alphap01fig}) due
to pseudoscalar mesons is reduced compared 
with the leading order result. The pseudoscalars themselves
now give $\Gamma_{0+-}=$ 81 eV rather than 106 eV, 
as in section IV.

\begin{figure}[htbp]
\begin{center}
\epsfxsize = 6.5cm
\ \epsfbox{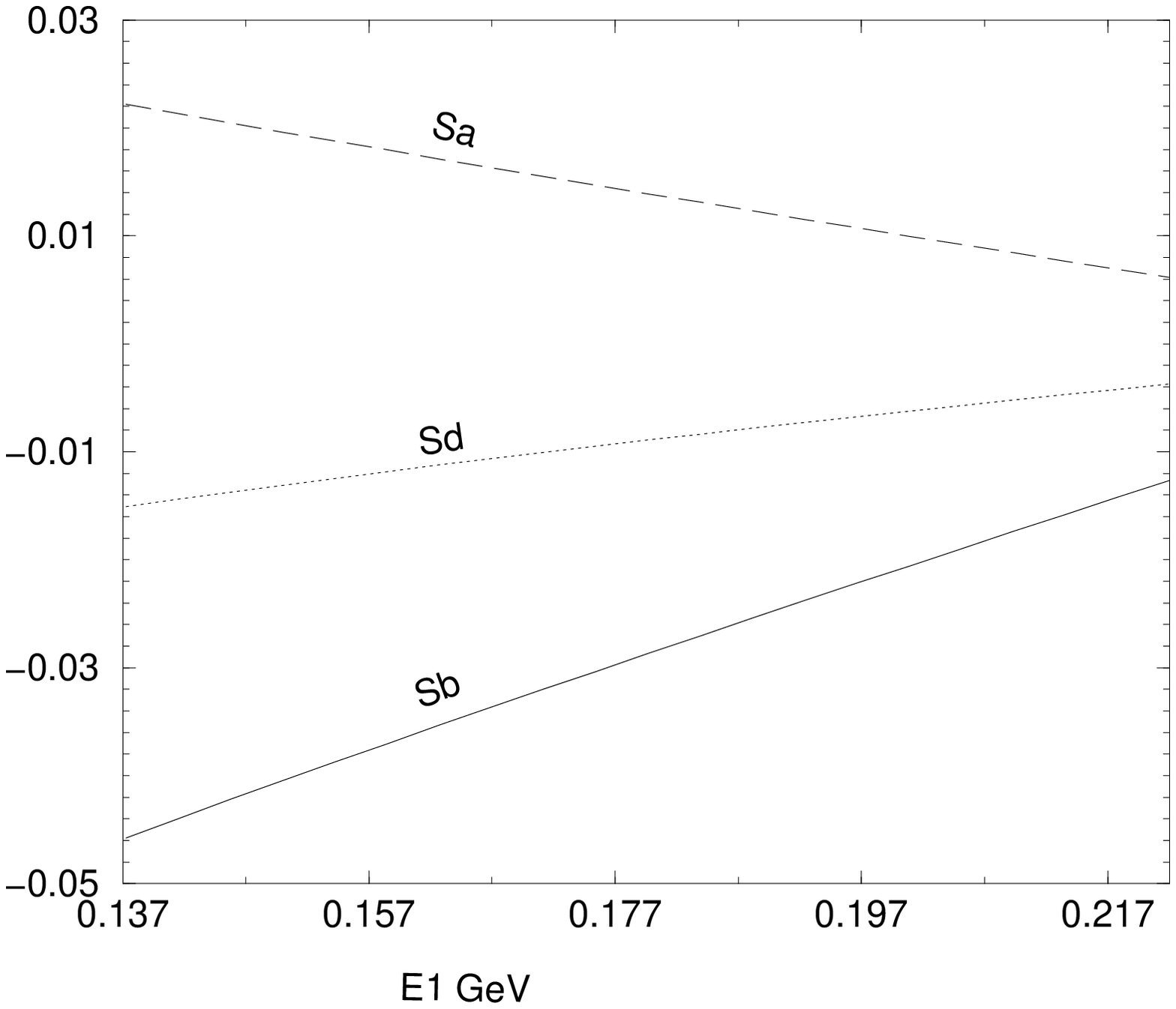}
\epsfxsize = 6.5cm
\ \epsfbox{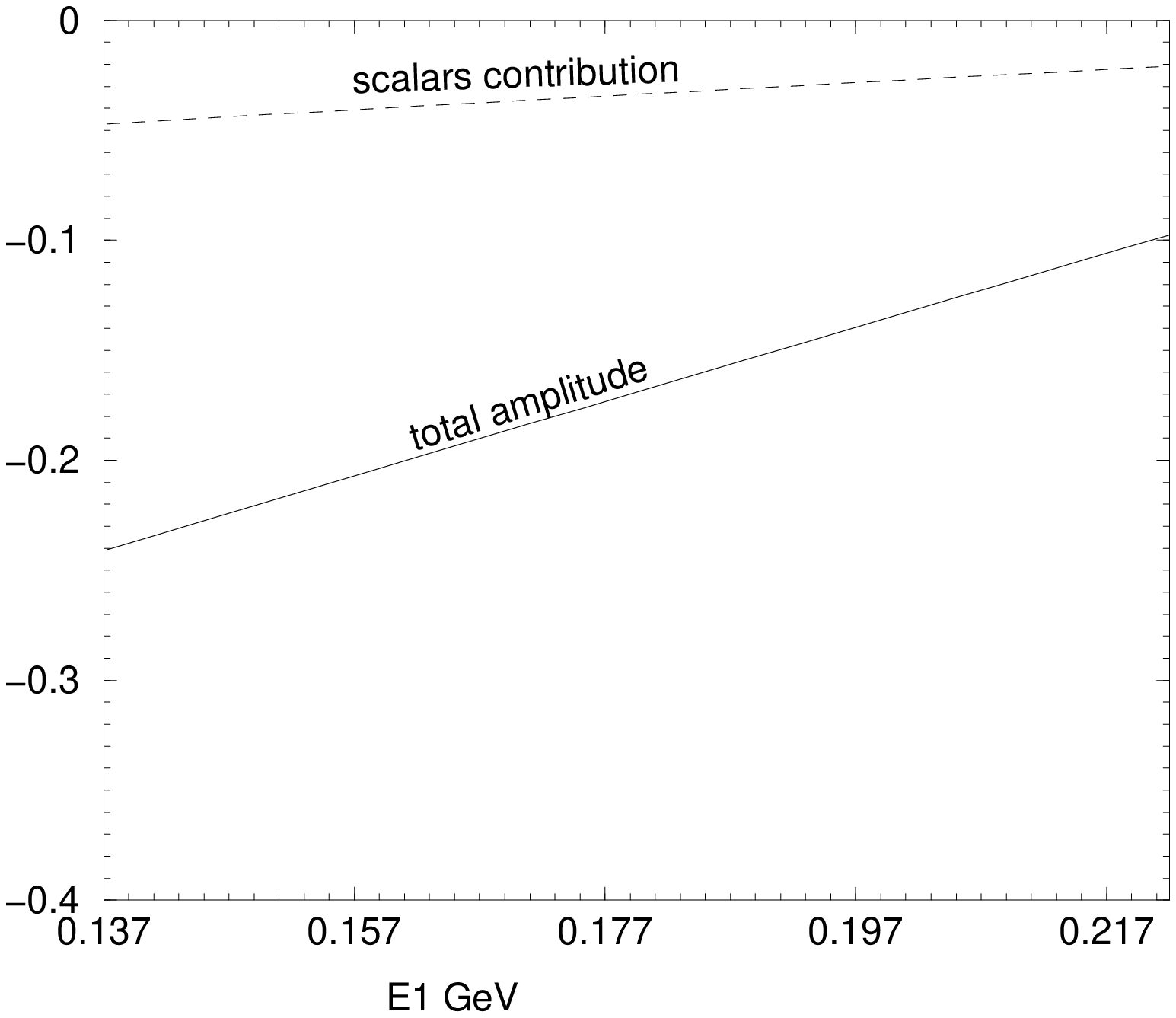}
\end{center}
\caption
{Left:  Contributions due to the three largest scalar 
meson diagrams, taking into account higher order symmetry 
breaking effects encoded in $\alpha_p$ and $\lambda^\prime$.  
Compare with the first of Figs.\ref{scalar_pseudoscalar01fig}.  
Right: Total amplitude due to pseudoscalar and scalar 
mesons, taking into account $\alpha_p$ and $\lambda^\prime$ 
terms (compare with second of 
Figs.\ref{scalar_pseudoscalar_alphap01fig}).}
\label{scalar_pseudoscalar_alphap03fig}
\end{figure}

However, for the diagrams involving scalar mesons the effect
of higher order symmetry breaking is even more important.
As we noted above, the scalar meson contribution
to $\eta \rightarrow \pi^+ \pi^- \pi^0 $ was 
rather small as the main diagrams tended
to cancel. When we include the $\alpha_p$ and
$\lambda^\prime$ corrections to the $\eta \pi^0$ transition
this cancellation will not be so complete.
Specifically, comparing Eqs.(\ref{etapitrans}) and
(\ref{newetapi}) we see that the magnitude of the 
amplitude for Fig.\ref{scdiagrams}a,
where the $\eta - \pi^0$ transition occurs with an on-shell
$\eta$, will be reduced by a factor of approximately
four, while that of 
Fig.\ref{scdiagrams}b, where the $\eta - \pi^0$
transition has an on-shell pion, will remain about the same
relative to our result in Section IV.  
Fig.\ref{scdiagrams}d will be unchanged. There will now 
be a non-negligible contribution from the scalar mesons. 
It will be more negative in 
sign (the contribution from Fig.\ref{scdiagrams}a is positive, 
but now smaller in magnitude)
and will add ``constructively'' to the pseudoscalar diagrams
in Fig.\ref{psdiagrams} and so increase our prediction for
$\Gamma(\eta \rightarrow \pi^0 \pi^+ \pi^-)$. This is shown
in Fig. \ref{scalar_pseudoscalar_alphap03fig}.    
Adding all of the pseudoscalar and scalar
diagrams in Fig.\ref{psdiagrams} and Fig.\ref{scdiagrams}
with the inclusion of the
symmetry breaking effects due 
to $\alpha_p$ and $\lambda^\prime$ we get:
$ \Gamma(\eta \rightarrow \pi^0 \pi^+ \pi^-)= 119.6 \, {\rm eV}$.
This is essentially the same as the result in section 
IV but now a larger portion is due to the scalar meson diagrams.  
Since this is the case the damping effect of the sigma width 
will be more prominent; in fact
it reduces the predicted rate to 103.6 eV in the 
Dashen's theorem limit.

It may be of interest to see how the detailed pattern just described  
depends on the precise value of the quark mass ratio $x$
and, as explained in section 4 of appendix A, 
correspondingly on the crucial isospin violating quark mass ratio 
$y$. This is shown for the predicted value of $\Gamma_{0+-}$
in Table \ref{firsttable}.

\begin{table}[htbp]
\begin{center}
\begin{tabular}{llll}
\hline
\hline
$x,y$ & ps. only & ps. + sc. (zero scalar widths) & 
ps.+sc. (non-zero scalar widths) \\
\hline
\hline
20.5, -0.202  & 63.7 eV  & 95.9 eV  & 82.2 eV \\
23, -0.241    & 70.7 eV  & 106.0 eV & 91.4 eV \\
25.1, -0.277  & 80.2 eV  & 119.6 eV & 103.6 eV \\
\hline
\hline
\end{tabular}
\end{center}
\caption[]{
$\Gamma(\eta \rightarrow \pi^+\pi^- \pi^0 )$ 
for different values of $x$ and $y$
defined after Eq.(\ref{spurion}). The second column 
applies to the case of only pseudoscalars present 
while the third includes scalars too. 
The effect of taking non zero scalar widths into 
account is shown in the last column. 
Dashen's theorem is assumed
in order to extract $y$.}
\label{firsttable}
\end{table}

\section{Including Vectors in the calculation}

It is well known that the inclusion of vector mesons is important for
a realistic discussion of low energy chiral dynamics. For example, in
the chiral pertubation scheme, most of the finite pieces of the
counterterms can be explained by integrating out various vector
contributions \cite{DRG}. In our present approach, of course, 
we are keeping the resonances, rather than integrating them out, 
in order to learn more about the scalars.  

First, the vector mesons contribute to the 
$\eta \rightarrow \pi^+ \pi^- \pi^0 $ amplitude
corresponding to Fig.\ref{vmdiagrams}a, which is just a 
correction to $\pi \pi$ scattering.
Its value given by the amplitude, $M_\rho^a$ 
in (\ref{vmampls}) is easily seen to be comparatively large.
However the fourth term of the $U(3)_L \times U(3)_R$
invariant Lagrangian Eq.(\ref{LagLO}) gives, in addition
to the $\rho \pi \pi$ vertex a four pion contact 
term [included in $M^a_{contact}$ in (\ref{psampls})]. 
Actually this contact term cancels most of the contribution 
from the $\rho$-exchange diagram in Fig.\ref{vmdiagrams}a.  
This is well known from chiral treatments of $\pi \pi$ 
scattering:  when the $\rho$ is added to the Lagrangian, 
chiral symmetry requires a contact term  which  cancels 
most of the $\rho$ contribution near threshold, thereby  
maintaining the current algebra threshold result.

However, the situation is actually a bit more complicated since  
the process of obtaining an adequate fit to the properties of both the
vectors and pseudoscalars \cite{Harada96} requires a number of
additional symmetry breaking terms shown in Eq.(\ref{LagSB}) 
of Appendix A. As well as the symmetry breaking terms we have 
already discussed involving the pseudoscalars alone, there are,
in particular, two new terms, measured by the coefficients 
$\alpha_+$ and $\alpha_-$. It turns out 
that their effects are very minor. They include
an additional contribution to the 4-point isospin violating 
$\eta \pi^+ \pi^- \pi^0$ vertex due to the
$\alpha_-$ term, corrections to the 4-pion vertex
in Fig.\ref{psdiagrams}b due to both $\alpha_+$ and $\alpha_-$  
and an additional diagram,
shown in Fig.\ref{vmdiagrams}b, which contains a 
new G-parity (and isospin) violating $\rho \pi \eta$ 
vertex [the amplitude for this is given
as $M_\rho^b$ in (\ref{vmampls})]. Note that
there there exists a $\rho^0 - \omega$ mixing
transition, which is the analog of the $\pi^0-\eta$ and
$a_0^0-f_0$ mixing transitions, but it does not 
contribute to \e3p at tree level. 

The decay widths with inclusion of vectors are tabulated in
Table \ref{secondtable} for the same values of $x,y$ used in the last 
section. In this table the neutral modes are also included. 
Furthermore,
the effect of both the scalar and (actually negligible) vector widths 
are included too. We see that, as expected from our discussion above,
the vectors do not change the overall predictions compared to the
last column of Table \ref{firsttable} very much but they do give a little 
enhancement. This is also clear by comparing the 
pseudoscalars + vectors column of table \ref{secondtable}
with the pseudoscalars only column of Table \ref{firsttable}. 
It is seen again that
the scalars make a non negligible contribution to the total amplitude.

\begin{table}[htbp]
\begin{center}
\begin{tabular}{lllll}
\hline
\hline
$x,y$ & decay mode & 
ps. +vec. &
ps.+sc.+vec.(no width)& 
ps. +sc. + vec. (width included)\\
\hline
\hline
20.5, -0.202 & 0+- & 64.4 eV &96.6  eV & 82.8 eV \\
             & 000 & 92.9 eV &139.4 eV &118.9 eV \\
\hline
23, -0.241 & 0+- & 71.9 eV   &107.4 eV & 92.5 eV \\
           & 000 & 101.9 eV  &152.9 eV &131.1 eV \\
\hline
25.1, -0.277 & 0+- & 82 eV &121.7 eV   & 105.4 eV \\
             & 000 & 114.5 eV &171.3 eV& 147.7 eV \\
\hline 
\hline
\end{tabular}
\end{center}
\caption[]{
$\Gamma (\eta \rightarrow \pi^0 \pi^+ \pi^-)$ 
and $\Gamma (\eta \rightarrow 3\pi^0)$ 
for different values of $x,y$. In the third column 
pseudoscalars and vectors are both present.
In the fourth and fifth column pseudoscalars, vectors 
and scalars all present (without and with the effect of 
the scalar meson widths).}
\label{secondtable}
\end{table}

Finally it is interesting to display the energy spectrum parameters $a,b$ 
and $c$ defined in Eq.(\ref{spectrum}) 
for the various models we have examined. 
These are given in Table \ref{energy_table} and are 
seen to be reasonable. The $\chi^2$ measures the fit of our 
model to the spectrum shape assumed in 
Eq.(\ref{spectrum}) and seems to be small.
\begin{table}[htbp]
\begin{center}
\begin{tabular}{lllll}
\hline
\hline
          &$a$    &$b$     & $c$    & $\chi^2$   \\
\hline
\hline
pseudoscalars(LO)    &-1.11  &0.31    & 0  &$5.6\times 10^{-5}$ \\
\hline
pseudoscalars        &-0.96  &0.23    &0   &$1.5\times 10^{-4}$ \\
\hline
pseudoscalars+scalars      &-0.93  &0.22 &-0.01 &$3.3\times 10^{-4}$ \\
\hline
pseudoscalars+scalars+vectors &-1.09  &0.26 &0.033   &$5.8\times 10^{-3}$ \\
\hline
\hline
\end{tabular}
\end{center}
\caption[]{Fits of the energy dependence of the normalized (charged)
decay amplitude for $\eta \rightarrow \pi^0 \pi^+ \pi^-$ 
to the form $|M_{0+-}|^2=1 +a Y + b Y^2 + c X^2 $.
The first line corresponds to result at leading order with pseudoscalar 
mesons only. The second with inclusion of higher order symmetry breakers,
the third when scalar mesons are added and the final line when vector
mesons are included as well.}
\label{energy_table}
\end{table}

\section{Discussion}

We studied the role of a possible nonet of light scalar mesons
in the still interesting $\eta \rightarrow 3\pi$ decay process.
Our motivation was primarily to learn more about the scalars
themselves. The framework is a conventional non-linear chiral
Lagrangian of pseudoscalars and vectors, extended to include scalars
(the Lagrangian is described in Appendix A). The parameters involving the 
scalars were previously obtained to fit the s-wave $\pi \pi$ and 
 $\pi K$ scattering in the region up to about 1 GeV as well
as the strong decay $\eta' \rightarrow \eta \pi \pi$. An initial concern
is whether the model as it stands, containing essentially no
undetermined main parameters (up to possible uncertainties in the quark
mass ratios $ x=2m_s/(m_u+m_d)$ and $y=(m_u-m_d)/(m_u+m_d)$), does
not make the $\eta \rightarrow 3 \pi$ amplitude too large.

In particular, the $\sigma(560)$ exchange diagrams (a) and (b) 
of Fig.\ref{scdiagrams} might lead to a great deal of enhancement due
to the possibility of the $\sigma(560)$'s momentum being close to mass 
shell. However this turns out not to be the case. In our initial
calculation where the scalars are added to the minimal model of 
pseudoscalars given in Eq.(\ref{lowestorderLag}), the left part
of Fig.\ref{scalar_pseudoscalar01fig} shows that these two diagrams,
though not individually small, tend to cancel each other. This
partial cancellation occurs because the $\eta-\pi^0$ transition leads
to opposite signs when the $\eta$ is on mass shell and when the
$\pi^0$ is on mass shell (In the first case we have a $\pi^0$
propagator carrying the momentum squared of an on-shell $\eta$
while in the second case, the reverse holds). In addition, the enhancement
due to the sigma propagator is further suppressed by the inclusion of an 
imaginary piece, needed to satisfy unitarity in the scattering calculation.
The net result is that the effect of including scalars in the minimal
pseudoscalar Lagrangian, Eq.(\ref{lowestorderLag}) is to increase the 
width for $\eta \rightarrow \pi^+ \pi^- \pi^0$ decay by about 13 per cent.
This relatively small, due to cancellation, increase illustrates the 
difficulty of finding  dramatic ``smoking gun" evidence for the existence 
of a light sigma. In the scattering calculation a light sigma appears (see 
for example \cite{HSS1}) obscured by a large background and does not have 
a simple Breit Wigner shape.

It is amusing to note the effect of higher derivative terms in the
Lagrangian of pseudoscalars (see section V for details). The higher 
derivative terms allow one to conveniently implement at tree level the 
fact that the ratio of the pseudoscalar decay constants $F_{Kp}/F_{\pi p}$
is somewhat greater than unity. With these terms the important $\pi^0 
\eta$ transition vertex has a momentum dependent piece. Together
with a needed modification in the parameter fitting (see section 4 
of Appendix A) this reduces the contribution of the pseudoscalars
to the $\eta \rightarrow 3\pi$ decay width. However the modification of 
the $\pi^0-\eta$ transition noticeably upsets the cancellation between the 
two sigma exchange diagrams in (a) and (b) of Fig.\ref{scdiagrams}.
The net result is that, while the total prediction for the $\eta 
\rightarrow 3\pi$ decay width remains about the same, now about
thirty percent of the value is contributed by the scalars.

The vector meson contribution, discussed in section VI, actually
does not change things much. This is because the $\rho$ exchange 
diagrams for $\pi\pi$ s-wave scattering are essentially canceled at very 
low energies by an extra four pion contact term which automatically arises 
due to the chiral symmetric formulation. Experimentalists fit
the Dalitz plot describing the $\eta \rightarrow \pi^+ \pi^-\pi^0$
spectrum to the form given in Eq.(\ref{spectrum}). A fit of this type
to the predicted spectrum from the Lagrangian of pseudoscalars,
scalars and vectors was seen to be close to the experimental one.
The basic spectrum shape is already reasonable with the very simplest 
model discussed in section III. As both the theory and experiment
get more precise, the importance of the spectrum shape toward
a deeper understanding of the underlying physics increases.

A particularly interesting scalar contribution to $\eta \rightarrow
 3\pi$ arises from the $a_0-\sigma$ transition shown in (d) of
Fig.\ref{scdiagrams} (The $a_0-f_0$ transition contribution to
$\eta \rightarrow 3\pi$ is suppressed due to the propagator of the heavier 
$f_0(980)$). This is the analog of the important $\pi^0-\eta$ transition
and, in a sense, is a new mechanism for $\eta \rightarrow 3\pi$
(although it was investigated a long time ago \cite{hudnall}
as a possible way to increase the $\eta \rightarrow 3\pi$ width).
The formula in raw form for this transition is given in 
Eq.(\ref{two-point-vertices}).
We evaluated its strength from the knowledge of the isospin violating 
piece of the dimensionless quark mass matrix ${\cal M}$ in 
Eq.(\ref{spurion}), determined from the pseudoscalar sector and the
coefficients: $a,b,c$ and $d$ of the scalar meson mass terms 
[see Eqs.(\ref{LagLO}] and (\ref{LagSB})) determined from 
the isospin conserving sector of the scalars. 
However as one can see from the left sides
of Figs.\ref{scalar_pseudoscalar01fig} and 
\ref{scalar_pseudoscalar_alphap03fig}, the contribution to $\eta 
\rightarrow 3\pi$ due to the $a_0-\sigma$ transition is not very large,
although it has the right sign to boost the decay rate.

The method just described also evaluates the strength of the 
$a_0(980)-f_0(980)$ transition. For convenience this is given after 
Eq.(\ref{scalar_isospin_violating_transitions}),
where the overall factor, $y$ is displayed. This transition has been very 
much ``in the news" recently as a proposed \cite{Close01} explanation for 
the large observed 
$\Gamma(\phi \rightarrow f_0\gamma)/\Gamma(\phi\rightarrow a_0\gamma)$ 
ratio and the anomalously strong $a_0$ central
production. However criticisms of this explanation have been given
\cite{Achasov02}, \cite{Black02}, pointing out that the $a_0-f_0$
mixing expected from a transition strength like the one determined above 
is insufficient to give a large effect. Intuitively, because of the near 
degeneracy of the $a_0(980)$ and $f_0(980)$ as well as the similarity
of their widths, one might expect the mixing to be very large. But
the mixing amplitude is governed by a dimensionless factor $iA_{af}/
(m_a\Gamma_a)$ [see for example Eq.(12) of \cite{Black02}] which
is suppressed by the scalar meson width, $\Gamma_a$.

In section II we discussed the current comparison between theory
and experiment for the $\eta \rightarrow \pi^0 \pi^+ \pi^-$ width. 
The experimental width \cite{pdg} is $\Gamma_{0+-}=267 \pm 25$ eV. 
This may be compared 
with the one loop chiral perturbation theory result \cite{Gasser85}
of $160 \pm 50$ eV. More recent attempts \cite{Kambor96} to estimate
final state interaction effect outside of the chiral perturbation theory
approach have increased this somewhat to $ 209 \pm 20$ eV. It seems 
to us that the thirty per cent contribution of the scalars compared to the 
pseudoscalars we have found should probably not be considered on top of 
this latter figure. That is because a good portion of the increase due to 
scalars we have found may be considered as resulting from final state 
interactions. Many attempts to close the gap between theory and 
experiment have focused \cite{Donoghue92} on a reanalysis of 
electromagnetic corrections to the $K^+-K^0$ mass difference. This is 
argued to increase the quark mass ratio, $y$ which is an overall factor for 
the $\eta \rightarrow 3\pi$ amplitude.

From the standpoint of learning more about the properties of the scalar
mesons it is clear that the $\eta' \rightarrow 3\pi$ decays represent
a potentially important source of information. In this case there is
sufficient energy available for the $a_0(980)$ and $f_0(980)$
propagators to be close enough to their mass shells to avoid suppressing
the contributions of these resonances. On the other hand, the
theoretical analysis is more difficult since large non-perturbative 
unitarity corrections are expected. In addition, other more massive 
particles may also contribute. The experimental information [see
Eq.(\ref{etap_experimental_rate})] 
is more preliminary than in the $\eta \rightarrow 3\pi$ case.
While a number with reasonably small errors has been presented for 
$\Gamma_{000}^\prime$, there is only a weak upper bound for 
$\Gamma_{+-0}^\prime$ and also no information on its Dalitz plot.
In the model employed in the present papper the $\eta' \rightarrow 3\pi$
amplitudes are simply obtained from the $\eta \rightarrow 3\pi$
amplitudes by the simple substitution given in 
Eq.(\ref{etap-amplitude}). 
Notice that this substitution rule would get modified if a more 
complicated $\eta-\eta'$ mixing scheme 
[e.g. the one mentioned after Eq.(\ref{eta_etap})]  
is adopted. As shown in Eq.(\ref{etap_LO_result}) 
the prediction of the minimal model of 
only pseudoscalars is somewhat too high, but at least of the correct order 
of magnitude. Adding the scalars without any readjustment of parameters
does not improve the prediction for $\Gamma'_{000}$ but makes it 
considerably larger (about 2300 eV). A similar large value was recently
found in \cite{paula}. Since the phase space is fairly large it
is perhaps to be expected that large values are typically obtained.
Presumably it is a sign for including more detailed unitarity corrections
or other physical effects which result in cancellations. We are 
particularly hopeful that a careful study of mixing between a lower mass
exotic scalar nonet and a more conventional higher mass scalar nonet
\cite{BFS3,Deirdreetal,2nonetmixing} may solve this problem and perhaps 
contribute to an improved understanding of the $\eta \rightarrow 3\pi$
decays also.

\acknowledgments
 We are grateful to Masayasu Harada, Paula Herrera-Siklody and Francesco
Sannino for very helpful discussions related to this problem.
The work of A. A-R. and J.S. has been supported in part by the US DOE
under contract DE-FG-02-85ER 40231.  D.B. wishes to acknowledge 
support from the Thomas Jefferson National Accelerator Facility 
operated by the Southeastern Universities Research Association (SURA) 
under DOE Contract No. DE-AC05-84ER40150. 
The work of A.H.F. has been supported by grants from the State 
of New York/UUP Professional Development Committee, and the 2002 
Faculty Grant from the School of Arts and Sciences, SUNY Institute 
of Technology. 

\begin{appendix}
\section{The chiral Lagrangian}

For convenience we collect here needed terms from the 
pseudoscalar-vector chiral Lagrangian presented in \cite{Harada96}
and from the scalar addition presented in \cite{BFSS2}.  
 
\subsection{Transformation Properties}

These are constructed to mock up the
symmetry properties of the fundamental quark Lagrangian, 
under which left and right projected light quark fields tranfsorm as
\begin{equation}
q_{L,R} \rightarrow U_{L,R} q_{L,R},
\end{equation}
$U_L$ and $U_R$ being $3 \times 3$ constant unitary matrices.  
The pseudoscalar nonet $\phi (x)$ is a $3 \times 3$ matrix which 
fits into the unitary chiral matrix 
\begin{equation}
U = {\rm exp} ( \frac {2 i \phi (x)}{F_\pi} )
\end{equation}
where $F_\pi$ is the (bare) pion decay constant.  Under a chiral 
transformation \begin{equation}
U \rightarrow U_L U U_R^\dagger.
\end{equation}
It is convenient to define the $3 \times 3$ unitary 
matrix $\xi$ by $U= \xi^2$.  Then $\xi$ transforms as 
\begin{equation}
\xi \rightarrow U_L \xi K^\dagger(\phi,x) = K(\phi,x) \xi U_R^\dagger,
\end{equation}
which implicitly defines the unitary matrix $K$.  
The intuitive significance of $K$ is that the objects $Kq$ behave 
like bare quarks surrounded by a pseudoscalar meson cloud, 
or ``constituent quarks''.  The objects
\begin{equation}
v_\mu \, p_\mu 
= \frac{i}{2} 
(\xi \partial_\mu \xi^\dagger \pm \xi^\dagger \partial_\mu \xi ),
\end{equation}
transform as 
\begin{eqnarray}
p_\mu &\rightarrow& K p_\mu K^\dagger \nonumber \\
v_\mu &\rightarrow& K v_\mu K^\dagger + i K \partial_\mu K^\dagger.
\end{eqnarray}
A putative scalar nonet matrix $N(x)$ is taken to transform as 
\begin{equation}
N \rightarrow K N K^\dagger,
\end{equation}
The vector meson nonet $\rho_\mu$ transforms as
\begin{equation}
\rho_\mu \rightarrow K\rho_\mu K^\dagger +
\frac{i}{\tilde g} K\partial_\mu K^\dagger,
\end{equation}
where we have included the dimensionless coupling constant, $\tilde
g$.  The ``field-strength tensor'' 
\begin{equation}
F_{\mu \nu} (\rho) = \partial_\mu \rho_\nu - \partial_\nu \rho_\mu - i
{\tilde g} \left[ \rho_\mu , \rho_\nu \right] \rightarrow K F_{\mu
\nu} K^\dagger.
\end{equation}

\subsection{$U(3)_L \times U(3)_R$ Invariant Terms}

These comprise the kinetic terms for the three multiplets, mass terms
for the scalars and vectors and appropriate interaction terms: 
\begin{eqnarray}
{\cal L}_0 = &-& \frac{F_\pi^2}{8} {\rm Tr} (\partial_\mu U \partial_\mu
U^\dagger) - \frac{1}{4} {\rm Tr} ( F_{\mu \nu}(\rho) F_{\mu
\nu}(\rho) ) \nonumber \\
& - & \frac{1}{2} {\rm Tr} ( {\cal D}_\mu N {\cal D}_\mu N ) -
\frac{m_v^2}{2 {\tilde g}^2} {\rm Tr} \left[ { ( \tilde g \rho_\mu -
v_\mu)}^2 \right] \nonumber \\
&-& a {\rm Tr} (NN) - c {\rm Tr} (N) {\rm Tr}(N) \nonumber \\
&+& F_\pi^2  \left[ A \epsilon^{abc} \epsilon_{def} N_a^d
{(p_\mu)}_b^e {(p_\mu)}_c^f + B {\rm Tr} (N) {\rm Tr}{(p_\mu p_\mu)} + C
{\rm Tr}(Np_\mu){\rm Tr}(p_\mu) + D {\rm Tr} (N) {\rm Tr} (p_\mu)
{\rm Tr} (p_\mu) \right],
\label{LagLO}
\end{eqnarray}
where ${\cal D}_\mu N = \partial_\mu N - i v_\mu N + i Nv_\mu$.  These
include the parameters $m_v^2, a, c, A, B, C$ and $D$. Note that the
pseudoscalars are still massless at this level. Further note that for
the interactions and mass terms of the scalars we do not restrict
ourselves to a single trace. For $q \bar q$ mesons the single trace
is suggested by the OZI rule while for an ideal dual nonet the
A term is in fact expected to be dominant.
We made a fit for $m_v^2, a, c, A, B, C$ and $D$
assuming only SU(3) invariance.

\subsection{Symmetry Breaking Terms}

The fundamental QCD Lagrangian contains the quark mass term $ - \frac{ (m_u +
m_d)}{2} \bar q {\cal M} q$, with the dimensionless matrix 
\begin{equation}
{\cal M} = \left[ \begin{array}{c c c}
1+y & 0 & 0 \\
0 & 1-y & 0 \\
0 & 0 & x 
\end{array} \right]
\label{spurion}
\end{equation}
where $x = \frac {2m_s}{m_u + m_d}$ and $y= - \frac{m_d - m_u}{m_d +
m_u}$.

It is convenient to define
\begin{equation}
{\hat {\cal M}}_{\pm} = \frac{1}{2} \left( \xi {\cal M} \xi \pm
\xi^\dagger {\cal M} \xi^\dagger \right).
\end{equation}

Then, the symmetry breaking Lagrangian is taken as
\begin{eqnarray}
{\cal L}_{SB} &=& \delta^\prime {\rm Tr} \left[ {\cal M} ( U + U^\dagger
) \right] + {\lambda^\prime}^2 {\rm Tr} \left[ {\cal M} U^\dagger
{\cal M} U^\dagger + {\cal M} U {\cal M} U \right] \nonumber \\
&-& \frac{2\alpha_p}{{\tilde g}^2} {\rm Tr} \left( {\hat {\cal M}}_+
p_\mu p_\mu \right) + 2 \alpha_+ {\rm Tr} \left[ {\hat {\cal M}}_+
\left( \rho_\mu - \frac {v_\mu}{\tilde g} \right) \left(  \rho_\mu -
\frac {v_\mu}{\tilde g} \right) \right] \nonumber \\
&-& \frac{2 \alpha_-} 
{\rm Tr} \left({\hat {\cal M}}_- \left[  \left(  \rho_\mu -
\frac {v_\mu}{\tilde g} \right), p_\mu \right]\right) \nonumber \\
&+& 2 \gamma^\prime {\rm Tr} \left[  {\hat {\cal M}}_+ F_{\mu \nu}
(\rho) F_{\mu \nu}(\rho) \right] \nonumber \\
&-& b {\rm Tr}(NN{\cal M}) - d {\rm Tr}(N) {\rm Tr}(N{\cal M}).  
\label{LagSB}
\end{eqnarray}

Only the parameters $\delta^\prime$, ${\lambda^\prime}^2$, $\alpha_p$,
$b$ and $d$ here contribute to the isospin violating vertices of
interest in the present paper. The parameter $\gamma^\prime$ was
included in the overall parameter fit obtained in \cite{Harada96} but
its small effect on the isospin-conserving vertices will be
neglected. 

In addition to the quark mass induced symmetry breaking terms there is
an important term induced by instanton effects which breaks just the
${U(1)}_A$ piece of  ${\rm SU}(3)_L \times {\rm SU}(3)_R 
\times {\rm U}(1)_V \times
{\rm U}(1)_A$. It may be summarized as
\begin{equation}
{\cal L}_{\eta^\prime} = \frac{\kappa}{576} {\rm ln}^2 \left( \frac
{ {\rm det} U}{ {\rm det} U^\dagger}\right) + \ldots
\label{extraetaprime}
\end{equation}
where $\kappa$ is a constant essentially proportional to the 
squared mass of the
$\eta^\prime$ meson.  The three dots stand for other terms which will
be neglected here but are listed in Eq. (2.12) of \cite{Schechter93}.
Effectively this term gives an important contribution to the
$\eta^\prime$ mass and an $\eta - \eta^\prime$ mixing angle defined by 
\begin{equation}
\left( 
\begin{array}{c} 
        \eta\\ 
        \eta' 
\end{array} 
\right) =
\left( 
\begin{array}{c c} 
{\rm cos} \theta_p  & -{\rm sin} \theta_p \\
{\rm sin} \theta_p  &  {\rm cos} \theta_p 
\end{array} 
\right)
\left( 
\begin{array}{c} 
 (\phi^1_1+\phi^2_2)/ \sqrt{2} \\ \phi^3_3 
\end{array} 
\right).
\label{eta_etap}
\end{equation}
When the extra terms in Eq. \ref{extraetaprime} are included they will not
only give rise to an additional isospin violating transition but will also
modify the $\eta-\eta'$ mixing transformation above to be the non-orthogonal
one given in Eq. (4.9) of \cite{Schechter93}. 
We will not include these effects in the
present paper, however.

\subsection{Numerical values of parameters used}
For the averaged pseudoscalar masses we used,
\begin{equation}
m_\pi = 0.137 \hskip.2cm {\rm GeV},
\quad\quad m_K = 0.4957 \hskip.2cm {\rm GeV}.
\end{equation}
In section III we gave the fitted parameters for the lowest order 
Lagrangian containing only pseudoscalars. This also yields the isospin
conserving quark mass ratio $x=25.1$ (assuming that $f$, defined
in Eq. (\ref{definef}) is unity). A refitting of these parameters
is necessary when the $\alpha_p/{\tilde g}^2$ and $\lambda'^2$
symmetry breaking terms are included. This can be conveniently done
 following the method used 
in preparing Table III of \cite{Harada96}. There, a value of $x$ is assumed
and the four quantities $F_\pi$ (unrenormalized pion decay constant),
$\delta'$, $|\lambda'|^2$ and $\alpha_p/{\tilde g}^2$ are calculated
in terms of the four physical quantities $m_\pi$, $m_K$, $F_{\pi p}=0.1307$
GeV and $F_{Kp}=0.1598$ GeV, using:
\begin{eqnarray}
\lambda'^2 &=& \frac{(1+x)F_{\pi p}^2m_\pi^2/16-F_{Kp}^2m_K^2/8}{1-x^2},
\nonumber \\
\delta'&=& F_{\pi p}^2m_\pi^2/8-4\lambda'^2,
\nonumber \\
\frac{\alpha_p}{{\tilde g}^2F_\pi^2} &=& \frac{(F_{Kp}/F_{\pi p})^2-1}{
2(1+x)-4(F_{Kp}/F_{\pi p})^2},
\nonumber \\
F_\pi &=& \frac{F_{\pi p}}{(1+4\alpha_p/({\tilde g}^2F_\pi^2))^{1/2}},
\nonumber \\
\frac{\alpha_p}{{\tilde g}^2} &=& F_\pi^2(\frac{\alpha_p}{{\tilde 
g}^2F_\pi^2}).
\label{paramfitting}
\end{eqnarray}
In addition, the isospin violating quark mass ratio $y$ is obtained from
\begin{equation}
(m_{K^0}^2-m_{K^+}^2)-f(m_{\pi^0}^2-m_{\pi^+}^2)=
(4y/F_{Kp}^2)(-2\delta'-8(1+x)\lambda'^2+m_K^2\alpha_p/{\tilde g}^2),
\label{gety}
\end{equation}
for a particular value of $f$. To isolate the effects of the scalars
we may choose an $x$ such that, with the value $f=1$
corresponding to Dashen theorem, we recover the value 
$y=-0.277$
found in sections III and IV. That gives $x=25.1$ and
\begin{eqnarray}
F_\pi = 0.128 \hskip.2cm {\rm GeV},
\quad\quad \delta'= 0.0386 \times 10^{-3} \hskip.2cm {\rm GeV}^4,
\nonumber \\
\alpha_p/{\tilde g}^2= 0.176 \times 10^{-3} \hskip.2cm {\rm GeV}^2, 
\quad\quad 
|\lambda'|=0.643 \times 10^{-3} \hskip.2cm {\rm GeV}^2.
\label{symbreakingparameters}
\end{eqnarray}

The needed dependences on the quark mass ratio ,$x$ of the parameters
involving vector mesons ($\gamma'$ , $\alpha_+$, $\alpha_-$, $m_v$
and $\tilde g$)
are given in Table 3 of \cite{Harada96}; the additional point $x=25.1$ used
in Table \ref{secondtable} above was treated by interpolation. 

The masses and widths of the scalars are taken to be (in MeV)
\begin{eqnarray}
m_\sigma = 550, \quad m_\kappa = 897, \quad m_{a_0} = 983.5, \quad
m_{f_0} = 980 \nonumber \\
\Gamma_\sigma = 370 , \quad \Gamma_{a_0} = 70.0, \quad \Gamma_{f_0}
= 64.6 .
\end{eqnarray}
Note that the values of $\Gamma_\sigma$ and $\Gamma_\kappa$ are not
``Breit-Wigner'' widths but are chosen to unitarize the $\pi \pi$ and
$\pi K$ scattering amplitudes.  The masses above fix the parameters
(in ${\rm GeV}^2)$ in (\ref{LagLO}) and (\ref{LagSB})
\begin{equation}
a = 0.492, \, b = -0.00834, \, c = -0.0160, \, d=-0.00557
\label{abcd-parameters}
\end{equation}
and the mixing angle $\theta_s = -20.3^0$.  The parameters $A,B,C,D$
define all the trilinear $S\phi \phi$ coupling constants according to
the formulas given in Appendix C of \cite{BFSS2}.  The needed
coupling constants are (in ${\rm GeV}^{-1}$)
\begin{eqnarray}
\gamma_{\sigma \pi \pi} = 7.27, \, \gamma_{\sigma \eta \eta} = 3.90,
\, \gamma_{\sigma \eta \eta^\prime} = 1.25, \, \gamma_{\sigma
\eta^\prime \eta^\prime} = -3.82, \nonumber \\
\gamma_{f\pi \pi} = 1.47, \, \gamma_{f \eta \eta} = 1.50, \,
\gamma_{f\eta \eta^\prime} = -10.19, \, \gamma_{f\eta^\prime
\eta^\prime} = 1.04, \nonumber \\
\gamma_{a\pi \eta} = -6.87, \, \gamma_{a \pi \eta^\prime} = -8.02.
\label{scalar_vertices}
\end{eqnarray}

\section{Decay amplitude}

The Feynman diagrams representing the $\eta(p) \rightarrow \pi^0 (p_1)
\pi^+ (p_2) \pi^- (p_3)$
decay are shown in Figs.1-3.  The contact diagrams (1a, 1b and 1c)  
receive contributions from the pseudoscalar and vector part of the
Lagrangian

\begin{eqnarray}
M_{contact}^a &=& 
i { { 16 y \delta' {\rm cos}\theta_p }\over  { 3  F_{\pi p}^4} } 
+ i { {  8 y \alpha_p {\rm cos}\theta_p }\over{3 {\tilde g}^2  F_{\pi 
p}^4} }
\left( - 3 p_2.p_3 + p.p_1 + p.p_2 + p.p_3  \right) 
+ i {  {512 y \lambda'^2 {\rm cos}\theta_p }\over  { 3 F_{\pi p}^4} }
\nonumber \\ 
M_{contact}^b &=& 
+ i \left( 
1 -  {  {3 m_v^2} \over {4 {\tilde g}^2  F_{\pi p}^2}   } 
   \right)
 { {C_{\pi \eta}^\eta}\over {m_\pi^2 - m_\eta^2} } 
 {2\over {3 F_{\pi p}^2}}
 \left( - 2 p_2.p_3 + p_1.p_3 - p.p_3 + p_1.p_2 - p.p_2 + 2p.p_1 
 \right)
\nonumber \\
&+&i\frac{2\alpha_+C_{\pi\eta}^\eta}{{\tilde g}^2F_{\pi p}^4(({m_\pi}^2-
{m_\eta}^2)}(-2p.p_1+2p_2.p_3+p_3.p-p_3.p_1+p_3.p-p_2.p_1)
\nonumber \\ 
&+& i { {8 \alpha_p\, C_{\pi\eta}^\eta}\over { 3  {\tilde g}^2  
F_{\pi p}^4 (m_\pi^2 - m_\eta^2) } }
 \left( 5 p.p_1  - 5 p_2.p_3 - p.p_3 + p_1.p_3 - p.p_2 + p_1.p_2 \right)
\nonumber \\ 
&+& i { {256 \lambda'^2 C_{\pi \eta}^\eta }\over
{ 3 F_{\pi p}^4 (m_\pi^2 - m_\eta^2 ) } }
+ i { {16 \delta'\, C_{\pi\eta}^\eta}\over { 3   
F_{\pi p}^4 (m_\pi^2 - m_\eta^2) } }
\nonumber \\ 
M_{contact}^c &=& 
i { {16 \delta' } \over {F_{\pi p}^4} }
\left(
{ { C_{\pi\eta}^\pi{\rm cos}^2 \theta_p}\over {m_\eta^2 - m_\pi^2 } } +
{ { C_{\pi\eta'}^\pi{\rm sin}\theta_p {\rm cos} \theta_p}
\over {m_{\eta'}^2 - m_\pi^2 } }
\right) \nonumber \\ 
&+& i { {8 \alpha_p}\over  { {\tilde g}^2  F_{\pi
p}^4} }
\left( 
{ { C_{\pi\eta}^\pi{\rm cos}^2 \theta_p}\over {m_\eta^2 - m_\pi^2 } } + 
{ { C_{\pi\eta'}^\pi{\rm sin}\theta_p {\rm cos} \theta_p}
\over {m_{\eta'}^2 - m_\pi^2 } }
\right)
\left( p.p_1 - p_2.p_3 + p.p_3 + p.p_2 -p_1.p_3 - p_1.p_2 
\right) \nonumber \\ 
&+& i { {256 \lambda'^2 }\over
{ F_{\pi
p}^4 } }
\left( 
{ { C_{\pi\eta}^\pi{\rm cos}^2 \theta_p}\over {m_\eta^2 - m_\pi^2 } } + 
{ { C_{\pi\eta'}^\pi{\rm sin}\theta_p {\rm cos} \theta_p}
\over {m_{\eta'}^2 - m_\pi^2 } }
\right)
\label{psampls}
\end{eqnarray}  

The scalar contributions (Figs.\ref{scdiagrams}a, b, c, and d) are:

\begin{eqnarray}
M_{scalar}^a &=& 
-i { {2 C_{\pi\eta}^\eta\gamma_{\sigma\pi\pi}^2}\over {m_\pi^2 - 
m_\eta^2}} { {(p.p_1)(p_2.p_3)} \over {m_\sigma^2 + (p-p_1)^2 }}
+ (\sigma \leftrightarrow f_0)
\nonumber \\
M_{scalar}^b &=& 
-i  \sqrt{2} 
\left( 
{ {2 C_{\pi\eta}^\pi \gamma_{\sigma\pi\pi} \gamma_{\sigma\eta\eta}}
\over {m_\eta^2 - m_\pi^2} } + 
{ { C_{\pi\eta'}^\pi \gamma_{\sigma\pi\pi} \gamma_{\sigma\eta\eta'}}
\over {m_{\eta'}^2 - m_\pi^2} }
\right)
{ {(p.p_1)(p_2.p_3)} \over {m_\sigma^2 + (p-p_1)^2 }}
+ (\sigma \leftrightarrow f_0)
\nonumber \\
M_{scalar}^c &=& 
-i   
\left( 
{ {C_{\pi\eta}^\pi \gamma_{a\pi\eta}^2}
\over {m_\eta^2 - m_\pi^2} } + 
{ { C_{\pi\eta'}^\pi \gamma_{a\pi\eta} \gamma_{a\pi\eta'}}
\over {m_{\eta'}^2 - m_\pi^2} }
\right)
{ {(p.p_3)(p_1.p_2)} \over {m_{a_0}^2 + (p-p_3)^2 }}
+ (p_2 \leftrightarrow p_3)
\nonumber \\
M_{scalar}^d &=& 
-i \sqrt{2} A_{a\sigma}\gamma_{a\pi\eta} \gamma_{\sigma\pi\pi}
{ {(p.p_1)(p_2.p_3)} \over 
{
  \left[{m_{a_0}^2 + (p-p_3)^2}\right]\left[{m_\sigma^2 + (p-p_1)^2 
}\right] 
} }
+ (\sigma \leftrightarrow f_0)
\label{scampls}
\end{eqnarray}

The $\rho$ contributions are:

\begin{eqnarray}
M_{\rho}^a &=&
 i 
{ 
  {m_\rho^4 C_{\pi\eta}^\eta }
             \over 
  { 2 {\tilde g}^2 F_{\pi p}^4 (m_\pi^2 - m_\eta^2) } 
}
{ 
  {p.p_1 + p_1.p_3 - p.p_2 - p_2.p_3 }
             \over 
  { m_\rho^2 + (p - p_3)^2 } 
}
+ (p_2 \leftrightarrow p_3) \nonumber \\
M_\rho^b &=& i \frac {4 \alpha_- y {\rm cos} \theta_p g_{\rho \pi
\pi}}{\tilde g F_{\pi p}^2} \frac{p_2 \cdot (p_3 - p_1)}{m_\rho^2 + {(p -
p_2)}^2} + ( p_2 \rightarrow p_3)
\label{vmampls}
\end{eqnarray}.

The two point vertices are:

\begin{eqnarray}
C_{\pi\eta}^\pi & = & - { {8 y \, {\rm cos} \theta_p \, \delta'}\over 
F_{\pi p}^2}
- { { 64 y \,\lambda'^2 \,{\rm cos} \theta_p }\over {F_{\pi p}^2 } } + 
{ {4 y \,\alpha_p \,{\rm cos}\theta_p \,m_\pi^2} \over {{\tilde g}^2 
F_{\pi p}^2 }}
\nonumber \\
C_{\pi\eta}^\eta & = & - { {8 y \,{\rm cos} \theta_p \,\delta'}\over 
F_{\pi p}^2}
- { { 64 y \,\lambda'^2 \,{\rm cos} \theta_p }\over {F_{\pi p}^2 } } + 
{ {4 y \alpha_p \,{\rm cos}\theta_p \,m_\eta^2} \over {{\tilde g}^2 
F_{\pi p}^2}}
\nonumber \\
C_{\pi\eta'}^\pi & = & - { {8 y \,{\rm sin} \theta_p \,\delta'}\over 
F_{\pi p}^2}
- { { 64 y \,\lambda'^2 \,{\rm sin} \theta_p }\over {F_{\pi p}^2 } } + 
{ {4 y \,\alpha_p {\rm sin}\theta_p \,m_\pi^2} \over {{\tilde g}^2 
F_{\pi p}^2}} 
\nonumber \\
C_{\pi\eta'}^{\eta^\prime} & = & - { {8 y \,{\rm sin} \theta_p \,\delta'}\over 
F_{\pi p}^2}
- { { 64 y \,\lambda'^2 \,{\rm sin} \theta_p }\over {F_{\pi p}^2 } } + 
{ {4 y \,\alpha_p {\rm sin}\theta_p \,m_{\eta^\prime}^2} \over
 {{\tilde g}^2 F_{\pi p}^2 
}}
\nonumber \\
A_{a\sigma} & = &
2 y\, ( b + d)\,  {\rm sin} \theta_s - \sqrt{2} y \, d \,{\rm cos}\theta_s 
\nonumber \\   
A_{af} & = &
- 2 y\, ( b + d)\,  {\rm cos} \theta_s - \sqrt{2} y\,  d \, {\rm 
sin}\theta_s 
\label{two-point-vertices}
\end{eqnarray}
Notice that the superscript on $C$ indicates which of the two
particles involved in the $\Delta I = 1$ transition is on-shell; this
only affects the $\alpha_p$ term which has derivative coupling.  

    It is not difficult to verify that the $\eta^\prime \rightarrow
\pi^0\pi^+\pi^-$ amplitude may be gotten from the one above by
simply making the interchanges
\begin{equation}
\eta \leftrightarrow \eta^\prime, {\rm cos} \theta_p \leftrightarrow 
{\rm sin} \theta_p,
\label{etap-amplitude}
\end{equation}
everywhere in Eqs. (B1)-(B3). This should not be done in Eqs. (B4)
since changing, for example, $C_{\pi\eta}^\pi$ to $C_{\pi\eta^\prime}^\pi$
accomplishes the desired result automatically.
\end{appendix}

\end{document}